\documentclass[11pt,a4paper]{article}

\usepackage[T1]{fontenc}
\usepackage[latin1,utf8]{inputenc}

\pdfoutput=1

\usepackage{amsmath,amssymb,amsfonts}
\usepackage[normal,font=small,labelfont=bf,labelsep=period]{caption}
\usepackage[pdftex]{color,graphicx}
\usepackage[english]{babel}
\usepackage[compress]{cite}

\usepackage{afterpage}

\usepackage[multiple]{footmisc}

\usepackage[dvipsnames]{xcolor}
\definecolor{darkgreen}{rgb}{0,0.65,0}
\usepackage{slashed}

\usepackage[linktoc=all,colorlinks=true,linkcolor=black,urlcolor=magenta,citecolor=black,hyperfootnotes=false]{hyperref}
\hypersetup{%
  colorlinks = true,
  linkcolor  = black
}

\newcommand{\mpl}{M_{\rm P}}
\newcommand{\dd}{{\rm d}}
\newcommand{\mS}{{\mathcal S}}
\newcommand{\mR}{{\mathcal R}}
\newcommand{\eq}[1]{(\ref{#1})}

\newcommand{\cs}{c_{\mathrm s}}
\newcommand{\taut}{\tilde{\tau}}
\newcommand{\zt}{z}

\def\bea{\begin{eqnarray}}
\def\eea{\end{eqnarray}}
\def\be{\begin{equation}}
\def\ee{\end{equation}}

\numberwithin{equation}{section}

\begin{document}

\begin{flushright}
\footnotesize
{IFT-UAM/CSIC-18-107}

\end{flushright}
\vspace{1cm}

\begin{center}
{\LARGE\color{black}\bf Black hole formation from a general quadratic action\\ for inflationary primordial fluctuations \\[1mm] }

\medskip
\bigskip\color{black}\vspace{0.6cm}

{
{\large\bf Guillermo Ballesteros,$^{1,2}$ Jose Beltr\'an Jim\'enez,$^{3}$ Mauro Pieroni$^{1,2}$}
}
\\[7mm]

{\it $^1$Instituto de F\'isica Te\'orica UAM/CSIC,\\
Calle Nicol\'as Cabrera 13-15, Cantoblanco E-28049 Madrid, Spain}\\
\vspace{0.2cm}
{\it $^2$Departamento de F\'isica Te\'orica, Universidad Aut\'onoma de Madrid (UAM)\\ Campus de Cantoblanco, 28049 Madrid, Spain}\\
\vspace{0.2cm}
{\it $^3$Departamento de F\'isica Fundamental and IUFFyM, Universidad de Salamanca,\\ E-37008 Salamanca, Spain}

\end{center}

\vspace{1cm}

\centerline{\large\bf Abstract}
\begin{quote}
\large 
The most up to date femto- and micro-lensing constraints indicate that primordial black holes of $\sim 10^{-16} M_\odot$ and $\sim 10^{-12} M_\odot$, respectively, may constitute a large fraction of the dark matter. We describe analytically and numerically the dynamics by which inflationary fluctuations featuring a time-varying propagation speed or an effective Planck mass can lead to abundant primordial black hole production. As an example, we provide an ad hoc DBI-like model. A very large primordial spectrum originating from a small speed of sound typically leads to strong coupling within the vanilla effective theory of inflationary perturbations. However, we point out that ghost inflation may be able to circumvent this problem. We consider as well black hole formation in solid inflation, for which, in addition to an analogous difficulty, we stress the importance of the reheating process. In addition, we review the basic formalism for the collapse of large radiation density  fluctuations, emphasizing the relevance of an adequate choice of gauge invariant variables. 
\end{quote}

\begin{center} 

\vfill\flushleft
\noindent\rule{6cm}{0.4pt}\\
{\small  \tt guillermo.ballesteros@uam.es, jose.beltran@usal.es, mauro.pieroni@uam.es}

\end{center}

\newpage

\tableofcontents

\section{Introduction}

The recent LIGO and VIRGO detections of binary black hole (BH) mergers in the approximate range of 7 to 36 $M_\odot$ \cite{Abbott:2016blz,Abbott:2016nmj,Abbott:2017vtc,Abbott:2017oio,Abbott:2017gyy}  have spurred a renewed interest on the idea that primordial black holes (PBHs) could contribute to the dark matter (DM) of the Universe \cite{Bird:2016dcv,Clesse:2016vqa,Sasaki:2016jop}. By PBHs we refer to BHs that have not formed as the endpoint of an astrophysical process (e.g.\ the collapse of a star) and that are old enough; in particular older than the epoch of nucleosynthesis; dating back to a redshift larger than $\sim10^9$. PBHs may have formed in the early Universe through a variety of mechanisms, such as the collapse of cosmic string loops \cite{Hawking:1987bn,Polnarev:1988dh}, bubble wall collisions from phase transitions \cite{Hawking:1982ga} and large density fluctuations \cite{1966AZh43758Z,Hawking:1971ei,Carr:1974nx}. In this work we will focus on the third possibility and, more concretely, on fluctuations originating on the dynamics of primordial inflation. If such perturbations are sufficiently large they can form BHs after inflation ends, when the comoving wavelength, $k$, that characterizes them becomes comparable to the Hubble scale. The mass of these BHs is inversely proportional to $k^2$ and, given that inflation is assumed to last at least $N\sim 50 - 60$ e-folds  and $k\sim \exp N$, the possible range of their masses is, a priori, huge. 

A lower limit on the mass of PBHs is imposed by Hawking radiation \cite{Hawking:1974rv,Hawking:1974sw} because PBHs lighter than $\sim 10^{-18} M_\odot$ would not survive until today. Moreover, the existence of a large population of PBHs heavier than {$\sim \mathcal{O}(10) M_\odot$ that could be of any relevance for the DM problem is excluded by several constraints; particularly their effects on the Cosmic Microwave Background (CMB) \cite{Blum:2016cjs, Ali-Haimoud:2016mbv,Poulin:2017bwe}, dwarf galaxies data \cite{Brandt:2016aco,Li:2016utv,Koushiappas:2017chw}, lensing of SNIa \cite{Zumalacarregui:2017qqd},\footnote{See however \cite{Garcia-Bellido:2017imq}.} observations in X-rays of our galaxy \cite{Gaggero:2016dpq}, wide binaries' orbital data \cite{2014ApJ...790..159M} and pulsar timing \cite{Schutz:2016khr}. Importantly, the BH merger rate estimates derived from the 
LIGO/Virgo observations \cite{Abbott:2017iws,Abbott:2016drs,Abbott:2016nhf} set --through the mechanism described in \cite{Nakamura:1997sm} (see also \cite{Ioka:1998nz})-- what at face value are the strongest constraints on the abundance of PBHs in that range \cite{Ali-Haimoud:2017rtz,Kavanagh:2018ggo,Chen:2018czv,Raidal:2017mfl,Ballesteros:2018swv}. This still leaves about 21 orders of magnitude in mass to explore in between the evaporation and stellar mass limits. In this broad range,} the abundance of PBHs is also constrained by different observations, notably microlensing studies by EROS/MACHO \cite{Tisserand:2006zx} and Subaru \cite{Niikura:2017zjd}. The range of masses that goes approximately from $10^{-16} M_\odot$ to $10^{-14} M_\odot$ was thought to be limited by femtolensing of gamma-ray bursts \cite{Barnacka:2012bm} until very recently. Interestingly, the actual constraining power of this data and part of the Subaru data (also relevant in that range) is currently under scrutiny due to several lensing effects that had not been properly accounted for in earlier works. In particular, it has recently been shown in \cite{Katz:2018zrn} that the range of PBH masses going approximately from $10^{-16}M_{\odot}$ to $10^{-14}M_\odot$ is currently unconstrained. It has also been pointed out in \cite{Inomata:2017vxo} that the range from $10^{-13} M_{\odot}$ to $10^{-10.5} M_{\odot}$, which could potentially be constrained by Subaru \cite{Niikura:2017zjd}, is  open as well because it requires a proper analysis of a so-called ``wave effect'', which is lacking at the moment.\footnote{Between $10^{-14} M_{\odot}$ and $10^{-13} M_{\odot}$, PBHs can account for at most $\mathcal{O}(10)\%$ of the DM, due to constraints from hypothetical encounters between PBHs and white dwarfs\cite{Graham:2015apa}.} The possibility exists that a very significant fraction of the DM ($\sim 100 \%$) could be constituted by PBHs in these mass ranges. See \cite{Ballesteros:2017fsr} for a concrete single-field model of inflation capable of producing a large abundance of PBHs in that range.

Beyond the motivation for PBHs as a DM candidate, much heavier $\mathcal{O}(10) M_\odot$ PBHs may also turn to be important for understanding current and near-future detections of binary mergers through gravitational waves. In particular, the identification of a merger component with a mass clearly below $1 M_\odot$ could hardly be explained with astrophysical formation mechanisms. Moreover, within the context of inflation, the detection of PBHs (or lack thereof) can be thought as a probe of very high energy physics operating during the early Universe. As inflationary subproducts, PBHs would be a window open today into the Universe at distance  scales that are inaccessible through the CMB and large scale structure (LSS) studies such as galaxy clustering. 

The part of the spectrum of primordial fluctuations that we have already measured with those probes has been found to be nearly scale invariant, Gaussian and adiabatic. Very importantly, it also features a tiny  $\mathcal{O}(10^{-9})$ amplitude, inferred from the CMB temperature anisotropies \cite{Smoot:1992td}. However, abundant PBH production from inflation is thought to require a significant enhancement of this amplitude (by several orders of magnitude, as we will discuss later). This implies that the approximate scale invariance of the spectrum at very large (CMB and LSS) scales needs to be broken at the smaller scales at which PBHs would form. In canonical single-field models of inflation this can occur if the inflationary potential has a near inflection point \cite{Garcia-Bellido:2017mdw,Kannike:2017bxn,Ezquiaga:2017fvi,Ballesteros:2017fsr,Hertzberg:2017dkh,Ozsoy:2018flq,Cicoli:2018asa,Dalianis:2018frf}, which quite generically alters the standard slow-roll dynamics \cite{Germani:2017bcs,Motohashi:2017kbs,Ballesteros:2017fsr} {(in \cite{Starobinsky:1992ts} it was already noticed that a singular point in the inflationary potential could enhance the spectrum of primordial perturbations)}. In spite of this, the basics of the enhancement mechanism can be qualitatively understood  noting that the primordial spectrum in the slow-roll approximation is, basically, inversely proportional to $\epsilon=-\dot H/H^2$, where $H$ is the Hubble expansion rate and $\dot H$ is its time derivative. For an approximately constant $H$, the function $\epsilon$ is proportional to $\dot\phi^2$, where $\phi$ is the background value of the inflaton field. Classically, an approximate inflection point --or a plateau that makes the potential flatter than at CMB scales, as it was already suggested in \cite{Ivanov:1994pa}-- makes the field slow down, enhancing the amplitude of the primordial spectrum at the scales associated to that part of the potential. 

In the {(vanilla)\footnote{We use ``vanilla'' to refer to the EFT of inflation presented in \cite{Cheung:2007st}, which assumes a single field, slow-roll and a possibly time-varying speed of sound. For simplicity we will often refer to it just as the EFT inflation, in particular in Section \ref{EFTPBH}.}} effective field theory (EFT) of inflation \cite{Cheung:2007st}, the slow-roll amplitude of the primordial spectrum depends not only on $\epsilon$ but also on the speed of propagation --or sound speed-- of the scalar fluctuations, $\cs$. For a field with a standard kinetic term, $\cs$ is equal to the speed of light ($\cs =1$), but it can be {significantly} smaller if higher order derivative terms are non-negligible.  A well-known example of a model that allows $\cs\neq 1$ is that of K--inflation \cite{ArmendarizPicon:1999rj,Garriga:1999vw}, where the action is some function of $\phi$ and $(\partial\phi)^2$ allowing a quasi-de Sitter phase. Generically, $\cs$ is a function of time. If its time variation, measured by $s=\dot \cs/(\cs H)$, is slow enough and if the standard slow-roll conditions hold, the primordial spectrum goes as $\sim H^2/(\cs\,\epsilon\,M_P^2)$, where $M_P$ is the (reduced) Planck mass. Therefore, a small value of $\cs$ may in principle have an analogous effect as a small $\epsilon$ and lead to PBH formation. This has recently been pointed out in \cite{Ozsoy:2018flq} and the  possibility of resonant production of PBHs from oscillations in $\cs$ has been studied in \cite{Cai:2018tuh}. 

A varying $\cs$  could lead to features on the primordial spectrum of particular relevance for the generation of PBHs. For instance, if the variation of $\cs$ and $\epsilon$ are temporarily synchronized, so that both decrease simultaneously over an interval, their combined effect in the power spectrum could allow to generate an enhanced narrow peak of fluctuations, possibly reducing the tuning that would be required to get the same peak from only one of the two functions.  Conversely, if $\cs$ and $\epsilon$ are diminished at significantly distant times during inflation, a primordial spectrum with two separate peaks would be created, leading to a bi-modal mass distribution of PBHs. If this happened $\sim 19$ e-folds apart, it could  account for a large fraction (or even the totality \cite{Katz:2018zrn}) of the DM with PBHs in the range  $10^{-16} M_\odot$ -- $10^{-14} M_\odot$ and, also, explain some of the binary mergers observed with $\mathcal{O}(30)M_\odot$ or smaller.

The EFT of inflation introduced in \cite{Cheung:2007st} does not contemplate the possibility of a time-varying effective Planck mass, $M\neq M_P$. Such a varying mass generically arises from non-minimal couplings, being the Brans-Dicke \cite{Brans:1961sx} type of theories a paradigmatic case where the scalar field is algebraically coupled. A more general framework is provided by the so-called Horndeski Lagrangians  \cite{Horndeski:1974wa,Deffayet:2011gz} which feature non-minimal derivative couplings and second order field equations despite having second order derivative terms in the Lagrangian, avoiding that way a potential Ostrogradsky instability \cite{Woodard:2015zca}. If the time variation of $M$ is sufficiently slow, the primordial spectrum for this kind of models goes as $\sim H^2/(\cs\,\epsilon\, M^2) $. The evolution of $M$ can thus lead to PBH production in a similar fashion to $\cs$ and $\epsilon$. In fact, $M$ and $\epsilon$ are indistinguishable at the level of the quadratic action for primordial fluctuations since they always enter through the combination $\epsilon M^2$. A specific model that explores this possibility to generate PBHs was already proposed in \cite{Suyama:2014vga}. This type of models of inflation is based on a tuned action (to ensure second order equations of motion) and cannot be easily justified in general neither as EFTs nor from an ultraviolet perspective.\footnote{See however \cite{Pirtskhalava:2015nla,Santoni:2018rrx}.} In spite of this, and due to its phenomenological interest, we delve in this work into the mechanisms through which the most general quadratic action for a single primordial fluctuation (with $\dot M\neq 0$) can lead to PBHs. 

Besides, we will show that the application of the EFT of inflation in the slow-roll regime suggests that quantitative estimates for {\it large} abundances of PBHs coming from a reduced $\cs$ are bound to be highly imprecise. The reason is that the interactions between primordial fluctuations grow as $\sim 1/\cs^2$ in the EFT of inflation. In consequence, a small $\cs$ has the effect of restricting the regime of validity of the theory, dooming its predictive power. Something analogous occurs as well in other EFT constructions, such as solid inflation \cite{Gruzinov:2004ty,Endlich:2012pz}. In fact, this ``fatal breakdown'' is expected to occur in any (plausible) EFT in which the interactions among fluctuations are dominated by the same control parameter, $\cs$, as the free part of the action. This is what happens in all examples known to us, which are based on background configurations that break one or several spacetime diffeomorphisms. Moreover, as the strength of self interactions among primordial fluctuations is parametrically enhanced by some positive power of $1/\cs$, non-Gaussianities typically increase for small values of $\cs$, contributing to a further degree of uncertainty on the predictions for PBH formation.

In the next section we review the basic formalism that is commonly used to estimate the abundance of PBHs from large inflationary perturbations. We revise the origin of the expression that relates the primordial perturbation to the total collapsing radiation density and point out a subtlety that becomes relevant in models such as solid inflation, where fluctuations can evolve outside the Hubble horizon. Then, in Section \ref{genactionM} we discuss how an enhanced primordial spectrum can arise from a generic quadratic action for a single independent primordial fluctuation. We provide a detailed analytical insight on the enhancement using the WKB approximation and extend previous works on the subject \cite{Leach:2000yw,Leach:2001zf}. We also give a couple of numerical examples for phenomenological parametrizations of the relevant functions of time ($\cs, \epsilon, M$). In Section \ref{EFTPBH} we discuss the difficulty that arises for PBH formation within the context of the EFT of inflation. In Section \ref{sec:single_field_models} we provide a toy model with $\dot \cs\neq 0$ which displays several of the features discussed in Section \ref{genactionM}. In Section \ref{pbhsolid} we discuss PBH formation in solid inflation, we explain why the reheating period after inflation is particularly important in this case and show, once more, the obstruction posed by the consistency of the (slow-roll) EFT if one demands a large abundance of PBHs. We present our conclusions in Section \ref{Conc}.

\section{PBH formation from large primordial fluctuations}

\label{largecollapse}

PBHs form after inflation ends whenever the  comoving wavenumber, $k$, of a sufficiently large density fluctuation becomes comparable to the Hubble scale, i.e.\ $k\simeq a\,H$, where $a$ is the scale factor of the Universe in a FLRW metric. The mass of the PBHs is approximately given by the energy density contained inside a Hubble patch at the time of collapse: $M\propto(4 \pi/3)\rho/H^3$, where $\rho$ is the average energy density of the Universe and the proportionality factor is $\lesssim 1$. For PBHs formed during radiation domination this gives (see e.g.\ \cite{Ballesteros:2017fsr} and references therein):
\begin{align} \label{Mk0}
M\simeq 10^{-15} M_\odot \left(\frac{k}{10^{14}\, \text{Mpc}^{-1}}\right)^{-2}\,.
\end{align}
{More precisely, we can write: 
\begin{align}\label{Mk}
M\simeq 10^{18}g \left(\frac{\gamma}{0.2}\right)\left(\frac{g(T_f)}{106.75}\right)^{-1/6} \left(\frac{k}{5\times 10^{13}\, \text{Mpc}^{-1}}\right)^{-2}\,,
\end{align}
where $\gamma$ is a factor that measures the efficiency of the process and $g(T_f)$ is the effective number of degrees for freedom in the radiation density at the time of formation.}

Given that $H$ is approximately constant during inflation, the expression \eq{Mk0} can be written in terms of $\Delta N^*$, the number of e-folds elapsed from the time at which the scale $k_*= 0.05$ Mpc$^{-1}$ (typically used to measure the CMB parameters) exits the horizon during inflation {\cite{Ballesteros:2017fsr}: 
\begin{align}
M\simeq M_\odot\,e^{36.74-2\Delta N^*}\,.
\end{align}

The abundance of PBHs of a given mass is usually computed using {an analogy} of the Press-Schechter formalism, according to which the mass fraction that collapses {into PBHs of mass equal or larger than $M$} is 
\begin{align} \label{GPS}
\beta(M)=(2\pi\sigma^2(M))^{-1/2}\int_{\delta_c}^{\infty} \textrm{d} \delta\, e^{-\delta^2/(2\sigma^2(M))}\,.
\end{align}
This expression makes two important assumptions: that the collapse occurs over a threshold $\delta_c$ and that the total density contrast $\delta$ is a Gaussian variable. If they hold true and $\delta_c$ is known, what ultimately determines $\beta(M)$ is $\sigma^2(M)$: the variance of the density contrast smoothed over a scale $R\sim 1/k$. Clearly, $\beta(M)$ is exponentially sensitive to $\sigma^2(M)$ and therefore the density contrast has to be very well-known to draw an accurate prediction from this expression. This implies that the spectrum of primordial perturbations at horizon entry also has to be known precisely. In radiation domination, the relation between $\delta$ and the comoving curvature perturbation, $\mathcal{R}$, that is commonly used in the literature to connect the PBH abundance to the spectrum of $\mathcal{R}$ is
\begin{align} \label{howeq}
\delta=\frac{4}{9}\left(\frac{k}{a H}\right)^2\mathcal R\,.
\end{align}
Then, the variance $\sigma^2(M)$ is
\begin{equation}\label{eq:beta1}
\sigma^2(M) =\frac{16}{81} \int \frac{  \textrm{d} q}{q} \left( q R \right)^4 \Delta_{\mathcal{R}}^2(q) W(q R)^2\,,
\end{equation}
where $\Delta_{\mathcal{R}}^2(q)$ is the dimensionless power spectrum of $\mathcal{R}$. For the smoothing window function, a Gaussian: $W(x)=\exp(-x^2/2)$, is usually chosen for convenience;  although other choices are possible and  $\beta(M)$ should (in principle) only be weakly dependent on them.\footnote{In practice, the result can depend on the window function, see \cite{Ando:2018qdb}.} The fraction of energy density on PBHs of mass equal or larger than $M$ relative to that of DM is:
\begin{align} \label{fDM}
\frac{\Omega_{PBH}}{\Omega_{DM}}\simeq \frac{\beta(M)}{6\times 10^{-16}} \left(\frac{M}{10^{-15}M_\odot}\right)^{-1/2}\left(\frac{\gamma}{0.2}\right)^{3/2}\left(\frac{g}{106.75}\right)^{-1/4}\,,
\end{align}
where $\gamma$ is an efficiency factor of the collapse process and $g$ is the effective number of degrees of freedom in the radiation density at formation time. The total energy density in PBHs is obtained integrating this expression.

{For the purpose of illustration, let us consider a contribution to the primordial power spectrum of the form $\Delta_\mathcal{R}^2(k_0)e^{-(k-k_0)^2/(2\Sigma^2)}$, representing a peak sufficiently separated from the CMB scales. In this expression $\Sigma$ is a constant that characterizes the width of the peak, whose height --given by $\Delta_\mathcal{R}^2(k_0)$-- is assumed to be much larger than the one measured by the CMB ($A_s\sim 10^{-9}$), so that for $k\sim k_0$ the spectrum $\Delta_\mathcal{R}^2(k)$ is given by the above expression to a very good degree of approximation. For any reasonable window function, it is a good approximation to write
\begin{align}
\sigma^2(M)\sim \Delta_\mathcal{R}^2(k) \simeq \Delta_\mathcal{R}^2(k_0)e^{-(k-k_0)^2/(2\Sigma^2)}\,,
\end{align}
and from here we obtain:
\begin{align}
\beta(M)\sim \frac{1}{2}\left(1-{\rm Erf}\left[y\,\frac{e^{(k-k_0)^2/(4\Sigma^2)}}{\Delta_\mathcal{R}(k_0)}\right]\right)\,,
\end{align}
where $y\sim O(0.1 -1)$ is a numerical factor that depends on the choice of $\delta_c$ --which is typically assumed to be $\delta_c\simeq 4/9$-- as well as on the shape of the window function. Using \eq{fDM} and \eq{Mk} and assuming $y\sim 1$ for the purpose of the argument, we can estimate  ${\Omega_{PBH}(M)}/{\Omega_{DM}}$ for any mass of interest. Focusing on the peak mass (i.e.\ $k=k_0$) and assuming $M(k_0)=M_\odot$, we find that ${\Omega_{PBH}(M)}\simeq{\Omega_{DM}}$ (for $M=M_\odot$) if $\Delta_\mathcal{R}^2(k_0)\simeq 0.0689$. If instead we choose $M(k_0)=10^{-15} M_\odot$, we require $\Delta_\mathcal{R}^2(k_0) \simeq 0.0318$ to have a peak abundance ${\Omega_{PBH}(M)}\simeq{\Omega_{DM}}$. The fraction of the DM in PBHs is very sensitive to the amplitude of the primordial power spectrum. For instance, in the previous examples, just reducing $\Delta_\mathcal{R}^2(k_0)$ to 0.0575 (for $M=M_\odot$) and 0.0294 ($M=10^{-15}M_\odot$) we obtain ${\Omega_{PBH}(M)}\simeq 0.05\, {\Omega_{DM}}$. Clearly, similar changes in $\delta_c$ (whose value is known only approximately) induce an analogous effect. The importance of these uncertainties (as well as that of others such as the window function) cannot be overemphasized enough. Although, they do not affect the prediction that PBHs will form from large primordial overdensities during radiation domination, they hamper the possibility of a very accurate prediction for the PBH abundance. This is even more so given that the formalism relies as well on other assumptions; for example that primordial fluctuations are Gaussian. An important take-home message of this discussion is that even barring all these uncertainties, small variations in the amplitude of primordial peaks have an exponential impact on the PBH abundance. This is why in this work we will be mostly concerned with the inflationary mechanisms that can generate such peaks, rather than with detailed estimates of the abundance.

Before moving on, let us finally notice that} for certain models (in which the primordial perturbations evolve outside the Hubble horizon) the relation \eq{howeq} should not be blindly applied and it is therefore worth explaining now in some detail its origin and its actual applicability. This will also serve us to address a couple of points  about the previous formulae that are seldom discussed in the literature on PBH formation from inflation.

\newcommand{\ch}{\mathcal{H}}
\newcommand{\rr}{\mathcal{R}}

\subsection{From primordial to radiation density fluctuations} \label{summa}
Let us consider a fluid with energy-momentum tensor $T_{\mu\nu}=(\rho+p) u_\mu u_\nu+p g_{\mu\nu}+\Pi_{\mu\nu}$ and (time-dependent) equation of state $w=p/\rho$. In the following we use primes to indicate derivatives with respect to conformal time, $\tau$, and we denote by $\ch=aH$ the conformal Hubble function. For convenience we use the (gauge invariant) Bardeen potentials $\Psi$ and $\Phi$ \cite{Bardeen:1980kt}, which can be easily identified in the {\it conformal Newtonian gauge} (CNG), which is defined by the choice of metric\footnote{We ignore vector (and tensor) fluctuations throughout.} 
\begin{align}
  \textrm{d} s^2 = a(\tau)^2[-(1+2\Psi)  \textrm{d} \tau^2+(1-2\Phi)\delta_{ij}   \textrm{d}x_i  \textrm{d} x_j].
\end{align}
For reasons that will become clear in Section \ref{whytmg}, we are interested in the {\it  total matter gauge} (TMG), whose linearly perturbed energy-momentum tensor can be easily written in terms of variables defined in the CNG \cite{Liddle:2000cg}: 
\begin{align} \label{rela1}
\delta_{M} & =  \delta_N+3(1+w)\ch v_N\\
\delta p_M & =  \delta p_N-p'v_N\,,\\
v_M & = v_N\,.
\end{align}
Here, the subscripts $_N$ and $_M$ refer to the CNG and the TMG, respectively. The velocity perturbation, $v$, is defined  in any gauge through $\delta T^i_0=-(\rho+p)\partial_i v$; and in the TMG corresponds to a gauge invariant quantity. The density perturbation in the TMG, $\delta_M=\delta\rho_M/\rho$, also corresponds to a gauge invariant quantity.\footnote{This gauge invariant quantity is denoted by $\epsilon_m$ in \cite{Bardeen:1980kt}.}

From now on we remove the subscript $_M$ from all TMG quantitities to simplify the notation.  In this gauge, the continuity and Euler equations in Fourier space are \cite{Liddle:2000cg}:\footnote{Note that \cite{Liddle:2000cg} uses dots to indicate derivatives with respect to conformal time.}
\begin{align} \label{cont}
\delta'-3w\ch\delta+(1+w)kV+2 w\ch \Pi=0\,,\\ \label{Euler}
V'+ \ch V = \frac{\delta P}{\rho}\frac{k}{1+w}-\frac{2}{3}\frac{w}{1+w}k\Pi+k\Psi\,,
\end{align}
where $\Pi_{ij}=\left(-{k_ik_j}/{k^2}+\delta_{ij}/3\right)\Pi$ and
\begin{align}
V=k\,v\,.
\end{align}  
In addition, $\delta$ and $\Pi$ satisfy the following Einstein equations: 
\begin{align} \label{const}
2k^2\Phi &= -3\ch^2 \delta\,,\\ k^2(\Phi-\Psi) & =3w\ch^2  \Pi\,. \label{const2}
\end{align}
The (gauge invariant) comoving curvature perturbation $\rr$ can be expressed  as follows:
\begin{align} \label{Req}
\mathcal{R}=-\Phi-{\mathcal H}v\,. 
\end{align}
If we replace $V$ and $\delta$ in \eq{cont} using, respectively, \eq{Req} and \eq{const} we get a differential equation for $\Phi$ with $\rr$ and $\Pi$ as sources:
\begin{align}\label{keyeq}
\frac{\Phi'}{\ch}+\frac{5+3w}{2}\Phi+\frac{3}{2}(1+w)\mathcal{R}-3\frac{\ch^2}{k^2}w\Pi=0\,.
\end{align}
If we now assume that the terms containing $\Pi$ and $\Phi' $ are negligible, we get a proportionality relation between $\Phi$ and $\mathcal R$. Then, using again \eq{const} we obtain 
\begin{align} \label{finala}
\delta\simeq\frac{2(1+w)}{5+3w}\frac{k^2}{\ch^2}\rr\,,
\end{align}
which reduces to \eq{howeq} for radiation domination ($w=1/3$).  For $k\ll \ch$, the time variation of $\rr$ vanishes for adiabatic perturbations (see e.g.\ \cite{Weinberg:2003sw}) and it is thus in this regime that \eq{keyeq} holds. 

PBHs form from the collapse of energy overdensities, which can only occur after Hubble crossing. It is then clear that the limit $k\ll \ch$ is not strictly valid. We will bear with the approximation of using \eq{keyeq}, but we point out that this prescription implies the extrapolation of a linear super-horizon relation to describe a non-linear process that occurs inside the horizon. 

\subsection{Why the total matter gauge?} \label{whytmg}

We have just discussed how the  comoving curvature perturbation, $\rr$, is linked to a gauge invariant variable, $\delta$, representing energy density fluctuations during radiation domination in the TMG. However, we have not explained why is this specific variable the one that should be considered for PBH formation. Indeed, we can define other gauge invariant quantities that can be identified with the energy density perturbation in other gauges different from the TMG. To show with an example why the variable $\delta=\delta_M$  is actually a good choice to describe PBH formation, let us compare it to the energy density in the CNG, $\delta_N$ (which can also be identified with a gauge invariant variable \cite{Bardeen:1980kt}).

The energy density is a Lorentz scalar, and therefore the change of its perturbation between any two gauges depends only on the time shift that relates them. The time shift that allows to go from the CNG to the TMG is given by the velocity perturbation; see \eq{rela1},  which defines the 	slicing of spacetime into hypersurfaces of constant time coordinate. Using the relation \eq{rela1} in \eq{finala} we obtain
\begin{align} \label{cont2}
\delta_N\simeq \frac{2(1+w)}{5+3w}\frac{k^2}{\ch^2}\rr-3(1+w)\frac{\ch}{k}V_N\,,
\end{align}
which is valid for $k\ll \ch$. During radiation domination and in the dominant adiabatic mode (see e.g.\ \cite{Ma:1995ey}) $V_N\simeq (k/\ch)\Phi/2\simeq -(k/\ch)\rr/3 $, where in the second equality we have used \eq{keyeq} and we are neglecting any possible source of anisotropic stress. Therefore, in {this regime} 
\begin{align}
\delta_N\simeq \frac{4}{3}\mathcal R +\ldots\,,
\end{align}
where the leading term in this expansion on $k/\ch$ comes from the contribution of $V_N$ to \eq{cont2}.

 Now we see that if we try to use naively $\delta_N$ instead of $\delta$ to compute the PBH abundance, the equation \eq{eq:beta1} becomes 
\begin{align}
\sigma^2_N(M)\simeq\frac{16}{9} \sigma^2_\mathcal{R}(M)\,,
\end{align}
which diverges due to the lower integration limit ($q\rightarrow 0$, where the primordial power spectrum is approximately scale invariant). This would imply $\beta(M)\rightarrow\infty$, regardless of the shape of the primordial spectrum at small scales. Notice in passing that if it were not because of the smoothing window function, the integral defining $\sigma(M)$ would also diverge (both for $\delta$ and $\delta_N$) due to its small scale (UV) behavior. 

This exercise exemplifies that an adequate definition of the quantity that we use to compute the abundance of PBHs is essential to make sense of the formalism summarized at the beginning of this Section.  Indeed, this is the nexus between the primordial spectrum of the comoving curvature perturbation and the PBHs. An appropriate variable playing this role should {satisfy} (at least) the following two requirements:
\begin{itemize}
\item Be gauge invariant.
\item Reduce to the Newtonian density perturbation for sufficiently small scales and low velocities.
\end{itemize}
Interestingly, it was shown in \cite{Chisari:2011iq} that the density perturbation that one can extract from Newtonian N-body simulations is precisely the one in the comoving orthogonal gauge (which is equal to the one in the TMG), i.e.\ $\delta$, and this identification is valid on scales comparable to the Hubble scale. As it was also argued by Bardeen \cite{Bardeen:1980kt}, $\delta$ is the most natural variable ``from the point of view of the matter'', because it corresponds to the energy density in the matter local rest frame. For all these reasons $\delta$ appears to be indeed a good variable to link PBHs to the primordial spectrum generated by inflation.

\subsection{Super-horizon primordial fluctuations}

So far, we have used all along the comoving curvature perturbation $\mathcal{R}$, defined in equation \eq{Req}, to connect the physics of inflation to the variable $\delta$ (the density contrast in the TMG) that is relevant to compute the abundance of primordial black holes. However, $\mathcal R$ is not the only gauge invariant variable that we can define and relate to primordial physics. Moreover, {other} such variables need not be equivalent to $\mathcal{R}$ in general. This {inequivalence can occur} under specific conditions (realized in concrete models) for which $\mathcal{R}$ is not conserved outside the horizon. This {may} introduce an ambiguity in the connection between $\delta$ and primordial physics that needs to be addressed in models  such as solid inflation \cite{Gruzinov:2004ty,Endlich:2012pz}, which is potentially interesting for PBH formation, as we will explore in Section \ref{pbhsolid}.

In addition to $\mathcal{R}$, it is common to use a gauge invariant variable, $\zeta$, that in the CNG reads
\begin{align} \label{defzeta}
\zeta=-\Phi+\frac{\delta_N}{3(1+w)}\,.
\end{align}
The relation between $\zeta$, $\mathcal{R}$ and $\delta\rho/\rho$ in the TMG is 
\begin{align} \label{xacte}
\zeta - \mathcal{R}=\frac{\delta}{3(1+w)}\,.
\end{align}
Since in this gauge $\delta = -2k^2\Phi/(3\ch^2)$, for $k$ sufficiently smaller than $\ch$ (i.e.\ for fluctuations whose comoving wavelength is much larger than the Hubble distance) if $\zeta$ is conserved $\mathcal R$ is also conserved (and conversely)  \cite{Weinberg:2008zzc}. Notice also that if the terms involving $\Phi' $ and $\Pi$ can be neglected in \eq{keyeq}, there is a simple approximate relation between $\Phi$ and $\mathcal{R}$, as we saw earlier, and then
\begin{align} \label{cruxeq}
\zeta \simeq \left(1+\frac{2}{3(5+3w)}\frac{k^2}{\mathcal{H}^2}\right)\mathcal{R}\simeq \mathcal{R}\quad \text{for}\quad k\ll\mathcal{H}.
\end{align} 
At the end of the day, we will be interested in computing the total density contrast using the (exact) expression \eq{xacte} at horizon re-entry (and during radiation domination), which is when PBHs start to form. If $\zeta$ and $\mathcal{R}$ are conserved for $k\ll \ch$, we can simply use \eq{cruxeq} into \eq{xacte}, which leads to \eq{finala} (and thus to \eq{howeq} for $w=1/3$). However, if $\zeta$ and $\mathcal{R}$ are not conserved we cannot use this simplification and we need to compute $\Delta_\delta^2$ by other means to feed it into 
\begin{align}
\sigma^2(M)=\int\frac{  \textrm{d} q}{q} \Delta_\delta^2(q) W(qR)^2\,,
\end{align}
which replaces \eq{eq:beta1} as the variance of the smoothed density contrast (which in turn goes into \eq{GPS} in order to get the PBH abundance). For instance, in Section \ref{pbhsolid} will make use of the basic assumptions made about reheating in solid inflation \cite{Endlich:2012pz} to get a straightforward expression for $\Delta_\delta^2$ from \eq{xacte} in that specific case. 

\section{Enhanced primordial power spectrum} \label{genactionM}

The power spectrum of scalar fluctuations generated during inflation is determined by their quadratic action around the (quasi) de Sitter background. If the inflationary model features only one independent scalar perturbation which becomes conserved on sufficiently large scales --and which we take in the following to be the curvature perturbation $\mR$, defined in \eq{Req}-- its most general quadratic action (at lowest order in derivatives) can be written as 
\begin{equation} 
\mS=  \int \dd t \, \dd^3 x \, M^2\frac{a^3 \epsilon}{\cs^2} \left[ \dot{\mR}^2 - \frac{\cs^2}{a^2 } \vert {\vec{\nabla}} \mR\vert^2 \right]\,.
\label{Eq:quadraticgeneral}
\end{equation}
In this expression dots indicate derivatives with respect to cosmic time $t$, $M(t)$ is an effective Planck mass for the fluctuations,  $\epsilon=-\dot H/H^2$ is the slow-roll parameter that guarantees inflation when $\epsilon <1$, the function $a(t)$ is the scale factor of the FLRW metric and $\cs(t)$ is the sound speed for the propagation of the primordial fluctuations $\mathcal{R}$. Clearly, the action \eq{Eq:quadraticgeneral} depends on just two independent combinations of these functions of $t$. In particular, $M^2$ and $\epsilon$ appear only multiplying each other and therefore are indistinguishable at the level of linear perturbations; however, it is important to notice that this degeneracy is broken by the background evolution so that it is convenient to  keep them both as it will become clear later.

Let us now comment on the types of field models that conform to the action \eq{Eq:quadraticgeneral}. If there is only one scalar field driving inflation, say $\phi$,  and no other fields are relevant while inflation lasts, there is a single independent scalar perturbation and its linear dynamics is captured by the action (\ref{Eq:quadraticgeneral}). If $\phi$ has a canonical kinetic term and rolls down a potential $V(\phi)$, then $\cs=1$, $M=\mpl$ and the linear dynamics of $\mathcal{R}$ is entirely determined by $\epsilon$, which is in turn determined by $V(\phi)$ (and the initial conditions for $\phi$ and $\dot\phi$). Assuming that the dynamics eventually reaches the slow-roll attractor \cite{Liddle:1994dx}, only the appearance of a special feature in the potential --such as an inflection point-- can give rise to the PBH production, as it has already been considered in \cite{Starobinsky:1992ts,Ivanov:1994pa,Garcia-Bellido:2017mdw,Kannike:2017bxn,Ezquiaga:2017fvi,Ballesteros:2017fsr,Hertzberg:2017dkh,Ozsoy:2018flq,Cicoli:2018asa,Dalianis:2018frf}. In order to have $\dot\cs\neq 0$,  a non-canonical kinetic term is required. The simplest example is that of models whose action is some function $p(X,\phi)$, where $X=-\partial_\mu\phi \partial^\mu \phi/2$. At the quadratic level and focusing on just two spatial derivatives acting on $\mathcal{R}$, these are the models contained in the EFT of inflation. As we already mentioned in the Introduction, the evolution of $\cs$ on time may generate PBHs, provided that the spectrum of $\mathcal{R}$ is sufficiently enhanced; see e.g. \cite{Ozsoy:2018flq,Cai:2018tuh}. 
 
A time-varying effective Planck mass can be generated via ``braiding'' (see \cite{Deffayet:2010qz}) or through a non-minimal coupling between the inflaton {and gravity}. However, not any coupling between the two is sufficient to have $\dot M\neq 0$. In particular, a coupling of the form $K(\phi) R$ can be recast (in the absence of other fields) into a redefinition of the scalar field potential $V(\phi)$. On the other hand, a coupling such as $K(X)R$ does in general lead to $\dot M\neq 0$, but at the expense of introducing an Ostrogradsky ghost. The latter can be avoided provided that this type of coupling appears with an appropriately weighted combination of higher derivative terms of $\phi$, see e.g. \cite{Deffayet:2009wt}. The general class of models constructed in this way that have second order equations of motion (and thus are free of instabilities at the classical level) is called after G.\ W.\ Horndeski, who described them for the first time in 1974 \cite{Horndeski:1974wa}. This class of models contains examples for which an effective $M(t)$ cannot be removed by any (conformal or disformal) field redefinition, except for some particular cases \cite{Bettoni:2013diz,Zumalacarregui:2013pma}. In  \cite{Suyama:2014vga} one such model was already proposed for PBH generation.  

All the examples we have mentioned so far are single-field models, but the action \eq{Eq:quadraticgeneral} can also capture the dynamics of primordial fluctuations for other models, e.g.\ some multi-field cases with distinctly heavy (and thus negligible) isocurvature perturbations (see e.g.\ the discussion in \cite{Turzynski:2014tza}) and, also, some models with massive vector fields as e.g.\ \cite{Jimenez:2016opp, DeFelice:2016yws}. Although, as we have seen, \eq{Eq:quadraticgeneral} applies to a wide variety of models, we can also entertain the possibility of going beyond it. First, we can consider including a mass term: 
\begin{equation} 
\mS=  \int \dd t \, \dd^3 x \, M^2\frac{a^3 \epsilon}{\cs^2} \left[ \dot{\mR}^2 - \frac{\cs^2}{a^2 } \vert {\vec{\nabla}} \mR\vert^2 - m^2\mR^2 \right]\,.
\label{Eq:quadraticgeneral2}
\end{equation}
where $m$ is yet another independent function of $t$. In Fourier space, the equation of motion for $\mathcal{R}$ becomes
\begin{align} \label{massiveR}
\ddot{\mathcal{R}}+\left(3-2\,s+\epsilon_2+\mu\right)H\dot{\mathcal{R}}+\left(\frac{\cs^2}{a^2}k^2+m^2\right) \mathcal{R}=0\,,
\end{align}
where we define $s=\dot \cs/(\cs H)$, $\epsilon_2=\dot\epsilon/(\epsilon  H)$ and $\mu=(M^2)^\cdot/(M^2 H)$. This equation clearly shows that the presence of the mass term prevents the existence of a conserved mode of the comoving curvature perturbation $\mathcal{R}$ for $\cs k\ll aH$. As we will see in Section~\ref{pbhsolid}, the equation for $\mathcal{R}$ in solid inflation \cite{Gruzinov:2004ty,Endlich:2012pz} is of the form \eq{massiveR}, with $m\sim \mathcal{O}(\epsilon) H$ being slow-roll suppressed. In addition, in this case $\mathcal{R}$ and $\zeta$ --which we defined in \eq{defzeta}-- are not equal to each other for small $k$ and therefore this model is outside the realm of those with a single independent scalar perturbation. It is worth mentioning as well that the non-conservation of $\mathcal{R}$ in the small $k$ limit also occurs, for instance, in multi-field models with a non-negligible entropy perturbation that originates in a curved trajectory in field space \cite{Gordon:2000hv}.

Another possibility that goes beyond \eq{Eq:quadraticgeneral} consists in including higher order derivative terms in the quadratic action for $\mathcal{R}$. As we already mentioned earlier, this generically introduces a ghost degree of freedom. This can be acceptable in the context of an EFT, as it is the case in the so-called ghost condensate \cite{ArkaniHamed:2003uz}, which we will briefly discuss in Section \ref{EFTPBH}. Cases like this one can be dealt with, to some extent, changing the dispersion relation for the Fourier modes of $\mathcal{R}$, including a momentum dependence on $\cs$. For instance, in the ghost condensate $\cs\sim k^2$ in the regime of validity of the EFT.

 In order to simplify our analysis, we will focus in this section on the action \eq{Eq:quadraticgeneral}, with $\cs$ and $M^2$ being functions of time only. As we have just summarized, this corresponds to the vast majority of the single-field models of inflation found in the literature. Let us then discuss how the functions $\epsilon$, $\cs$ and $M^2$ can give rise to an amplification of the primordial power spectrum that will lead to the generation of PBHs. 
 
 At this point, it is useful to absorb the effect of $\cs$ into a rescaled time variable $\taut$ that generalizes the usual conformal time $\tau$ as follows:
\be
\dd \taut=\cs\,\dd \tau=\frac{\cs}{a}\,\dd t.
\ee 
Doing so we effectively restore the equal scaling of time and space coordinates in the action, which now reads:
\begin{equation}
\mS= \frac{1}{2} \int \dd \taut \, \dd^3 x \, \zt^2 \left[ (\mR')^2 - \vert \vec{\nabla} \mR\vert^2 \right]\,,
\end{equation}
where primes stand for the rest of this section for derivatives with respect to $\taut$ and we have defined
\be
\zt^2\equiv \frac{2 M^2 a^2\epsilon}{\cs}.
\label{eq:defz}
\ee
Notice that, unlike in the usual case where $z$ is defined to be dimensionless, here we are defining $z$ including the effective mass Planck so that it has dimension of mass. The next step is to canonically normalize the scalar perturbation as 
\bea
v\equiv \,\zt\, \mR\,.
\eea 
We obtain 
\begin{equation}
\mS= \frac{1}{2} \int \dd \taut \, \dd^3 x \left[ (v')^2 - \vert \vec{\nabla} v\vert^2 +\frac{\zt ''}{\zt} v^2\right] \,,
\end{equation}
which leads to the (generalized) Mukhanov-Sasaki equation that, in Fourier space, reads 
\be
 v''+\left(k^2-\frac{\zt ''}{\zt}\right)v=0\,.
\label{eq:GMS}
\ee
One can then use the standard expressions for the solutions of this equation in the slow-roll regime, (remembering the previous definition of $\zt$ or, equivalently, the re-scaling of the time coordinate). Our interest here is to study the possibility of obtaining an enhancement of the power spectrum by means of a variation in the sound speed $\cs$ and/or the time-varying Planck mass $M^2$ in order to extend the more standard enhancement obtained from the time evolution of $\epsilon$ (as e.g.\ in single field inflation models featuring an inflection point in the potential). To illustrate the main idea  we will briefly review the computation of the power spectrum within this scenario. This will also lead us to obtain some relevant expressions that are not found in the literature. For convenience we define the following hierarchies of slow-roll functions:
\be
\epsilon_{i+1}\equiv{\epsilon_i}^{-1}\frac{\dd \epsilon_i}{ \dd N},\quad\quad s_{i+1}\equiv s_i^{-1}\frac{\dd s_i}{\dd N}\quad\quad{\rm and}\quad\quad\, \mu_{i+1}\equiv \mu_i^{-1}\frac{\dd \mu_i}{\dd N},
\ee
with $\epsilon_0=1/H$, $\epsilon_1=\epsilon$, $s_0=\cs$, $s=s_1$ and $\mu_0=M^2$. Here $N$ denotes the number of e-folds defined as
\be
\dd N=H\dd t=\frac{aH}{\cs}\dd\taut
\label{eq:defN}
\ee
and which, by convention, grows as inflation progresses forward in time. We will say that the {\it generalized slow-roll} regime holds provided $|\epsilon_i|\,,|s_i|, |\mu_i|\,\ll 1,\,\forall i\geq1$.

We start, as usual, by considering very short wavelengths $k^2\gg \zt''/\zt$ for which we have the same behaviour as in Minkowski spacetime so that, after appropriately normalizing and choosing the adiabatic Bunch-Davis vacuum, the solutions for the mode functions are
\be
v\simeq \frac{e^{-i k\taut}}{\sqrt{2k}}\,.
\label{eq:BDsolution}
\ee
On the other hand, long wavelength modes with $k^2\ll \zt''/\zt$ have solutions given by
\be \label{gensol}
v\simeq C_{1,k}\, \zt+C_{2,k}\,\zt\int\frac{\dd \taut}{\zt^2}\,.
\ee
If the second mode (determined by $C_{2,k}$) decays sufficiently fast (as guaranteed if we are in generalized slow-roll), the above solution quickly converges to $C_{1,k}\,z$ and the amplitude of $C_{1,k}$ can be obtained by matching with the short wavelength solutions:
\be
\vert C_{1,k}\vert^2=\frac{1}{2k \zt^2_*}
{=\frac{1}{2k}\left(\frac{\cs}{2M^2a^2\epsilon}\right)_*}
\ee
where the subscript $*$ stands for evaluation at the time $\taut_*$ where the matching is performed. The solution for very long wavelengths can then be written as
\be
\vert v\vert\simeq\frac{1}{\sqrt{2k}}\left\vert\frac{z}{z_*}\right\vert\,.
\label{eq:solvSH}
\ee
In order to compute the (best) matching time, we consider the first order correction to the frequency of the modes in the WKB approximation \cite{1926ZPhy...38..518W,1926ZPhy...39..828K,BrioullinWKB}, which is  given by
\be
\omega_k\simeq k\int\dd \taut\sqrt{1-\frac{\zt''}{k^2\zt}}\simeq k\int\dd \taut\left(1-\frac{\zt''}{2k^2\zt}\right)\,.
\ee
From this expression we see that the WKB regime will break down when the first order correction, which is implicitly assumed to be the dominant one,\footnote{More precisely, the WKB regime is valid as long as $|\omega^{-1}{\dd^n\omega}/{\dd\taut^n}|\ll (\mathcal{H}/\cs)^n$ for all $n$. Thus, it could happen that the first order correction remains small but the series breaks down due to some higher order derivative of $\omega$ becoming large. In principle, this might be expected to occur in some of the examples considered in this work, if the first slow-roll parameters remain small but higher order parameters could be large. However, we have obtained our results numerically solving the Mukhanov-Sasaki equation and compared them with the analytical approximations, finding and excellent agreement where it is expected. For a more detailed discussion on the WKB approximation and its conditions of validity within the context of particle production in cosmological scenarios see \cite{Winitzki:2005rw}.} becomes of order one,
so that the time $\taut_*$ of {\it horizon crossing}\,\footnote{The term ``horizon crossing'' is adopted here (and throughout the paper) by abuse of language, but the time $\taut_*$ does not, in general, correspond to the mode crossing any specific cosmological horizon.} that determines the matching is given for each $k$ by the solution of the equation
\be
k^2=\frac{\zt''}{2\zt}\Bigg\vert_{\taut_*}=\left[\frac{a^2H^2}{\cs^2}\big(1+\Xi\big)\right]_{\taut_*}\,,
\label{eq:crossing}
\ee
where in the second equality we have simply expanded the derivatives of $z$ in terms of the slow-roll parameters and defined the function 
\begin{align}
\Xi\equiv&
-\frac12\left(1+\frac{\epsilon_2}{2}-\frac{s}{2}+\frac{\mu_1}{2}\right)\epsilon
-\frac12\left(\frac{5}{2}+\epsilon_2-\frac{3}{4}s+\frac{s_2}{2}+\mu_1\right)s
+\frac14\left(3+\frac{\epsilon_2}{2}+\epsilon_3\right)\epsilon_2\,\nonumber\\
&+\frac{1}{4}\left(3+\epsilon_2+\frac12\mu_1+\mu_2 \right)\mu_1.
\end{align}

Equipped with the solution for the modes on very large scales together with the equation for the matching time, we can now write the following expression for the primordial power spectrum of $\mR$:
\be
\Delta_\mR^2=\frac{k^3}{2 \pi^2}\vert \mR_k\vert^2=\frac{k^3}{2 \pi^2}\vert C_{1,k}\vert^2=\frac{1}{8\pi^2}\left[\frac{(1+\Xi)H^2}{\epsilon\,\cs M^2}\right]_*\,.
\label{eq:PSsr}
\ee
This clearly shows that, assuming $H'/M_P\simeq 0$, a decrease in the slow-roll parameter $\epsilon$ will induce a peak in the primordial power spectrum, potentially leading to PBHs. We can also see that the same effect can be achieved from a decrease in the sound speed $\cs$ or the effective Planck mass $M^2$. The extra freedom at our disposal can either enhance the amplification in the power spectrum (if $\cs$, $M^2$
 and $\epsilon$ present a decrease at similar times, which is rather reasonable if the desired features originate from the same underlying mechanism) or several peaks at different scales, thus producing an approximately multi-modal distribution of PBHs, as discussed in the Introduction.

It is worth noting that $\Xi\ll 1$ applies in the generalized slow-roll regime, where we recover the usual sound horizon crossing condition $k=aH/\cs$. Obviously, $\Xi\ll 1$ also occurs if only the specific combination of slow-roll parameters appearing in $\Xi$ is small. In particular, this is the case if $\epsilon_{1,2,3}$ and $s_{1,2}$ remain small, while higher order slow-roll parameters could be large. Indeed, it is important to notice that since the enhancement in the amplitude of the power spectrum occurs by a non-negligible variation of $\epsilon_1$ or $\cs$, this can be accompanied by large values of some higher order slow-roll parameters, which may imply a breakdown of the generalized slow-roll regime. It is thus instructive to recall now some of the assumptions made to arrive at  \eq{eq:PSsr} in order to understand better its regime of validity.

\subsection{Multiple crossings}

An assumption implicit in \eq{eq:PSsr} is that there is only one instance of horizon crossing. However, there may be situations (of specific interest for us concerning the generation of PBHs) where a given band of modes can experience multiple crossings. This will be reflected in the existence of more than one solution to (\ref{eq:crossing}), which determines the crossing times. If this is the case, we can nonetheless straightforwardly extend the expression (\ref{eq:PSsr}), under additional (but reasonable) assumptions. Let us consider in particular that a certain mode undergoes three crossings so that it exits the horizon at some time $\taut_1$, re-enters the horizon at $\taut_2$ and finally exits at $\taut_3$. We are going to assume that outside the horizon, the solution is always quickly attracted to $v\propto \zt$. We will come back to this assumption in more detail in Sections \ref{Sec:Decaying} and \ref{Sec:General}. After the first crossing, the mode evolves as in the usual case with only one crossing, so the solution is
\be
\vert v^I(\tilde\tau)\vert^2 \simeq 
\frac{1}{2k}\left\vert\frac{\zt}{\zt_1}\right\vert^2,
\label{Eq:matchtau1}
\ee
where we have introduced the super-script $I$ to denote that it corresponds to the solution after the first crossing. The second crossing corresponds to the mode transiting from super- to sub-horizon, so the mode is inside the horizon again and the solution will be a general plane wave of the form
\be
v^{II}\simeq D_{1,k} e^{-ik\taut}+D_{2,k} e^{ik\taut}.
\ee
Unlike in the initial state, the mode after re-entering will not match a Bunch-Davis state, but some excited state with a linear combination of positive and negative frequency modes described by $D_{1,k}$ and $D_{2,k}$ (i.e.\ there will be some non-trivial Bogoliubov coefficients). These constants can be obtained by matching this solution to (\ref{Eq:matchtau1}), but we will not need their specific forms for our purposes. The amplitude of the mode during this phase is thus
\be
\vert v^{II}\vert ^2=\vert D_{1,k}\vert ^2+\vert D_{2,k}\vert ^2+2\vert D_{1,k}\vert\vert D_{2,k}\vert\cos\left(2k\taut+\delta_k\right)
\ee
where $\delta_k=\delta_{k,2}-\delta_{k,1}$, with $\delta_{k,1}$ and $\delta_{k,2}$ are the phases of $D_{k,1}$ and $D_{k,2}$, respectively. The matching at $\taut_2$ with the solution outside the horizon (\ref{Eq:matchtau1}) gives 
\be
\vert v^{II}(\taut_2)\vert ^2=\frac{1}{2k}\left(\frac{\zt_2}{\zt_1}\right)^2.
\ee
Finally, the mode evolving according to $v^{II}$ will re-exit the horizon at $\taut_3$ and will match the solution $v^{III}\simeq E_k z$, with some constant  $E_k$. The final amplitude of the mode can thus be computed by matching this solution and $v^{II}$ at $\taut_3$. The resulting expression can be related to the previous evolution of the mode as
\begin{align}
\vert v^{III}(\taut_3)\vert ^2=&\left\vert\frac{v^{III}_k(\tilde \tau_3)}{v^{II}_k(\tilde \tau_2)}\right\vert^2\left\vert\frac{v^{II}_k(\tilde \tau_2)}{v^{I}_k(\tilde 	\tau_1)}\right\vert^2 \vert v^{I}(\taut_1)\vert ^2
=\frac{1}{2k}\left(\frac{\zt_2}{\zt_1}\right)^2\left(\frac{1+\beta_k \cos\left(2k\taut_3+\delta_k\right)}{1+\beta_k\cos\left(2k\taut_2+\delta_k\right)}\right)^2\,,
\end{align}
where
\be
\beta_k\equiv\frac{2\vert D_{1,k}\vert\vert D_{2,k}\vert}{\vert D_{1,k}\vert ^2+\vert D_{2,k}\vert ^2}.
\ee
Since $\vert E_k\vert =\vert v^{III}(\taut_3)\vert/\zt_3$, the power spectrum then becomes
\be
\Delta_\mR^2=\frac{k^3}{2\pi^2}\vert E_k\vert^2=\frac{k^2}{4\pi^2}\left(\frac{\zt_2}{\zt_1
\zt_3}\right)^2\left[\frac{1+\beta_k \cos\left(2k\taut_3+\delta_k\right)}{1+\beta_k\cos\left(2k\taut_2+\delta_k\right)}\right]^2.
\label{Eq:Pkosc}
\ee
We thus see that the power spectrum receives an extra factor with respect to \eq{eq:PSsr} due to the extra time that it has remained sub-horizon between the second and third crossings and, furthermore, it can also exhibit some oscillations from the Bogoliubov coefficients that are generically produced in this process. It is not difficult (although tedious) to generalize this result to the case when there are more crossings. However, this is enough to show that it is expected to see some oscillations in the power spectrum in the range of $k$-modes that undergo a re-entering and re-exit, as we will see explicitly in the examples below.   {A modulated oscillation in the power spectrum was also found in \cite{Starobinsky:1992ts} in the context of single field inflation with singular points in the inflationary potential. In \cite{Polarski:1992dq}, analogous oscillations were shown to be generated in double field inflation with an intermediate matter phase. In this case, the mechanism for the generation of oscillations is similar to ours, i.e., the existence of a transient phase where the initial Bunch-Davis vacuum is excited to a non-vacuum state which leads to the oscillations in the power spectrum once it definitely leaves the horizon. It is important to notice however, that, while in \cite{Polarski:1992dq} the mechanism relies on an intermediate phase where inflation stops, the oscillations that we obtain here are genuinely produced by a modified dynamics of the perturbations and, as a matter of fact, our background can remain quasi-de Sitter from the beginning of inflation until the very end of it. 

Before proceeding to show explicit examples showing the oscillations in the power spectrum, let us go back to the assumption that the constant mode is quickly reached and analyze under which circumstances it holds. This will allow us to reveal another mechanism to enhance the power spectrum.
}

\subsection{The ``decaying'' mode}\label{Sec:Decaying}
An important assumption needed to arrive at the expressions \eq{eq:PSsr} and \eq{Eq:Pkosc} for the power spectrum of $\mR$ is that the second mode --see \eq{gensol}-- always decays faster than $z$ (which typically grows like $a$ in the strict generalized slow-roll regime) so that the mode $C_{1,k}\, z$ is reached quickly enough. If this is not the case, not only the above formulas for $\Delta_\mR^2$ cannot be applied, but also the presence of a second growing mode --which might grow faster than $z$-- can in turn induce an enhancement of the power spectrum by itself. In order to illustrate and understand these points, we will work with the equation for the curvature perturbation:
\be
\mR''_k+2\frac{\zt'}{\zt} \mR'_k+ k^2\mR=0\,.
\label{eq:Rintime}
\ee
Sometimes, it is also useful to write this equation in terms of the number of e-folds defined in (\ref{eq:defN}), and we give it here for completeness:
\be
\frac{\dd^2 \mR}{\dd N^2}+\Big(3-\epsilon+\epsilon_2-2s_1+\mu_1\Big)\frac{\dd \mR}{\dd N}+\frac{\cs^2k^2}{a^2 H^2}\mR=0.
\label{eq:Refolds}
\ee
From the definition of the function $z$ in (\ref{eq:defz}), it is easy to obtain
\be
\frac{\zt'}{\zt}=\frac{aH}{\cs}\left(1+\frac{\epsilon_2-s_1+\mu_1}{2}\right)\,.
\label{eq:friction}
\ee
This expression can be integrated so that $\zt$ satisfies
\be
\log\frac{\zt}{\zt_\star}=(N-N_\star)+\frac{1}{2}\int_{N_\star}^N\big(\epsilon_2-s_1+\mu_1\big) \dd N\,,
\ee
where $\star$ denotes some arbitrary time (not to be confused with the horizon crossing time denoted by $*$ above). We then see that $z\propto a$ in the strict slow-roll regime or, more generally, as long as the combination $(\epsilon_2-s_1+\mu_1)$ remains small. 
 
 For the very long wavelengths with sufficiently small $k$, the curvature perturbation equation reduces to
\be
\mR''_k+2\frac{\zt'}{\zt} \mR'_k\simeq0\,,
\ee
which, after one integration,  gives $\mR'\propto z^{-2}$ and the solution is
\be
\mR \simeq C_{1,k}+2\,C_{2,k}\int\frac{\dd \taut}{\zt^2}
=C_{1,k}+C_{2,k}\int  \frac{\cs^2}{a^3M^2\epsilon H} \dd N\,,
\label{eq:Generalsoloutside}
\ee
with $C_{1,k}$ and $C_{2,k}$ some integration constants which must be determined, as usual, from the chosen vacuum mode solutions. The first term is the well-known adiabatic mode that is conserved outside the horizon and that is guaranteed to exist due to zero-mode residual symmetries (see e.g. \cite{Weinberg:2003sw}). This is the mode upon which the results of the previous section are built upon. The second term $C_{2,k}$ is the usually decaying mode that typically becomes negligible within a few e-folds after horizon crossing so that the constant mode is quickly reached. In fact, the time $N_\star$ is intended to  represent some time soon after horizon crossing such that the decaying mode can be safely neglected and the constant mode dominates. This is what happens in the generalized slow-roll regime, where $\zt\propto a$ and, therefore, the second mode decays as $a^{-3}$. In this section we are interested in the situation where this mode is, on the contrary, the dominant one. In order to discern when this is the case, let us write the expression for the variation of $\mR$ with the number of e-folds $N$:
\bea
\frac{\dd \mR}{\dd N}=C_{2,k}\,e^{-3N}\frac{\cs^2}{M^2\epsilon\, H}
=C_{2,k}\exp\left[-\int \Big(3-\epsilon_1+\epsilon_2-2s_1+\mu_1\Big)\dd N\right]\,,
\label{eq:dRdN}
\eea
which clearly shows that the second mode decays as $a^{-3}\propto e^{-3N}$ in the generalized slow-roll regime. More precisely, the derivative of the curvature perturbation is exponentially suppressed as long as the combination 
\bea \label{frictious}
\xi\equiv3-\epsilon_1+\epsilon_2-2s_1+\mu_1
\eea
is positive. In that case, the constant mode $C_{1k}$ is quickly reached and the formulae for the power spectrum obtained above will be valid. If  the {\it friction parameter} $\xi$ is instead negative for some interval of e-folds, the derivative \eq{eq:dRdN} is no longer suppressed and the second mode can become the dominant one. Such a growing mode can also enhance the power spectrum and lead to PBH production. Interestingly, the time variation of the sound speed and the effective Planck mass contribute to $\xi$ through $s_1$ and $\mu_1$ and (from a model-building point of view) this freedom can either guarantee that the second mode never becomes the dominant one by appropriately keeping at bay any (positive) growth of $\epsilon_1-\epsilon_2$ or, instead, turn the second mode into a growing one, thus helping to enhance the power spectrum.

\subsection{General analysis} \label{Sec:General}
After reviewing the constant and decaying solutions for the super-horizon regime, we will give a more detailed discussion of the general solution. For that, we will start by rewriting the equation (\ref{eq:Rintime}) as
\be
\frac{1}{z^2}\frac{\dd}{\dd \taut}\left(z^2 \frac{\dd\mR}{\dd \taut}\right)=-k^2  \mR\, .
\ee
From this equation, it is easy to see that the exact solution for the curvature perturbation can be given in terms of the following integral equation
\be
\mR=\mR_\star\left[1+\frac{\mR'_\star}{\mR_\star}\int_{\taut_\star}^{\taut}\dd \taut_1\frac{z_\star^2}{z^2}-k^2\int_{\taut_\star}^{\taut}\frac{\dd \taut_1}{z^2}\int_{\taut_\star}^{\taut_1}z^2\frac{\mR}{\mR_\star}\dd \taut_2     \right]
\ee
where $\star$ indicates that the quantity it accompanies is evaluated at some arbitrary $\taut_\star$. Typically, one is interested in $\taut_\star$ corresponding to some time soon after horizon crossing. We can alternatively express all the quantities in terms of the number of e-folds as
\be
\mR=\mR_\star\left[1+\frac{1}{\mR_\star}\left(\frac{\dd \mR}{dN}\right)_\star\int_{N_\star}^{N}\dd N_1\frac{\bar{c}_{\rm s}}{\bar{a}\bar{H}\bar{z}^2}-\left(\frac{\cs k}{aH}\right)_\star^2\int_{N_\star}^{N}\dd N_1\frac{\bar{c}_{\rm s}}{\bar{a}\bar{H}\bar{z}^2}\int_{N_\star}^{N_1}\dd N_2     \frac{\bar{c}_{\rm s}\bar{z}^2}{\bar{a}\bar{H}}\frac{\mR}{\mR_\star}\right]
\label{eq:generalsolution}
\ee
where a bar over a quantity denotes that it is normalized to its value at $N_\star$. This expression extends the analogous result obtained in e.g. \cite{Leach:2001zf} to the more general scenario considered here with time-varying sound speed and effective Planck mass. In the above formula, we can identify the first two terms with the constant and ``decaying'' modes discussed in the precedent sections. The third piece can be used to obtain recursively the solution at the desired order in $k^2$. In the generalized slow-roll regime and well outside the horizon, the zeroth order solution is simply $\mR_\star$ which coincides with the asymptotic value of $\mR$ at the end of inflation.\footnote{Let us stress that this value of $\mR_\star$ can then be identified with the constant $\mR_*$ used in (\ref{eq:PSsr}) and whose value is obtained by matching with the sub-horizon solution at the horizon crossing defined by \eq{eq:crossing}. Although this gives the correct result for the super-horizon power spectrum, it is important to realize that $\mR_*$ will not coincide with the exact value of $\mR$ at horizon crossing (see e.g. the discussion in this respect in \cite{Kinney:2005vj}), but this is harmless since all that is required is that $\mR_*$ gives the correct asymptotic super-horizon value.} Interestingly, in that case, it is the $\mathcal{O}(k^2)$ correction coming from {the third} term what gives the leading order correction to the constant solution because it decays as $a^{-2}$, while the usual decaying mode goes as $a^{-3}$. This is important to properly identify each mode when computing higher order corrections to the constant solution.

If we abandon the usual slow-roll regime, there are two different mechanisms able to amplify the scalar power spectrum. On the one hand, if the friction term $\xi$ defined in (\ref{frictious}) becomes negative, then the usual decaying mode given by the second term inside the brackets in (\ref{eq:generalsolution}) turns into a growing mode which will then give rise to an enhancement of $\mR$. On the other hand, the leading order correction $\mathcal{O}(k^2)$ can also produce an enhancement of the curvature perturbation $\mR$ if the double integration in the third term inside the bracket in (\ref{eq:generalsolution}) gives a large enough contribution as to compensate for the suppression coming from $\left(\frac{\cs k}{aH}\right)_\star^2$. As argued in \cite{Leach:2001zf}, this might happen if $z^2$ becomes sufficiently small for some number of e-folds. In the most general case, the situation can be quite complex featuring a combination of all the effects we have discussed. However, in many physically relevant situations, the effects can be conveniently disentangled.

\afterpage{\clearpage} 
 \begin{figure}[t]
\includegraphics[width=8cm]{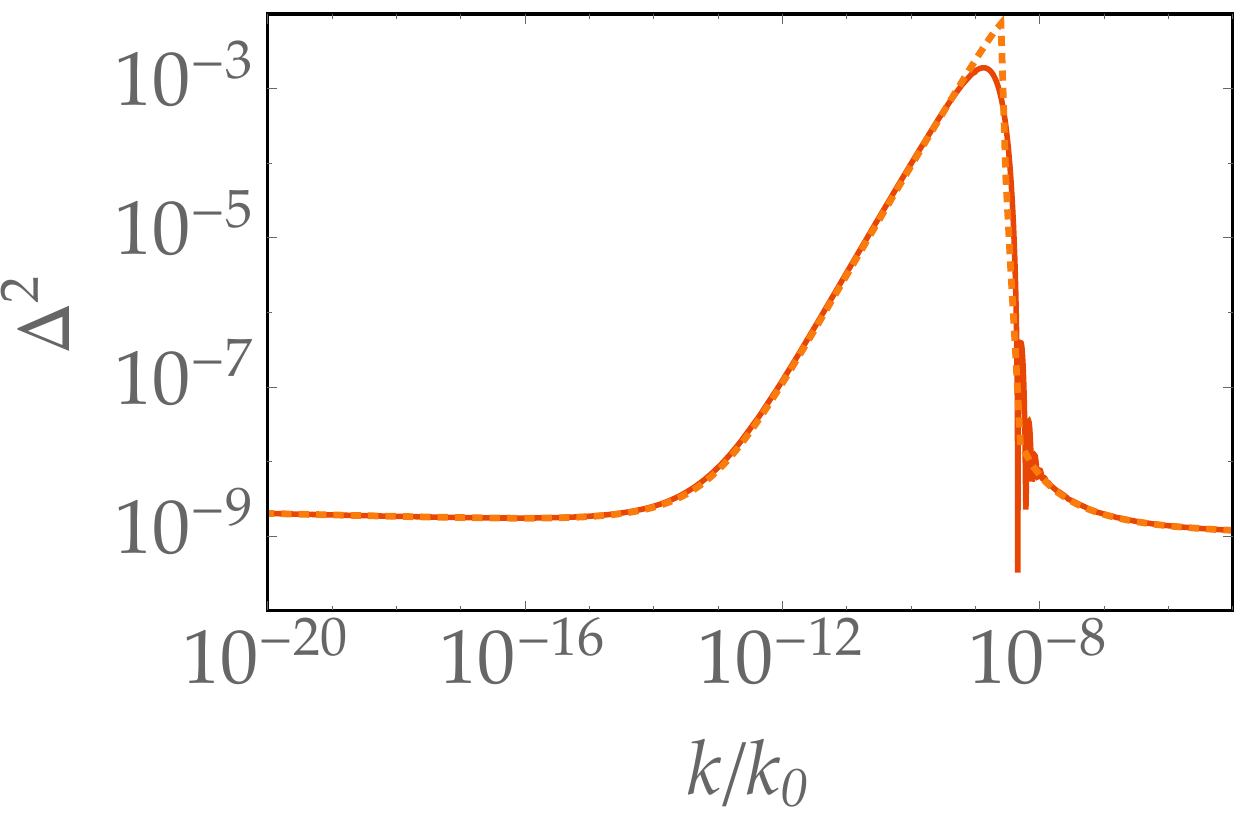} 
\includegraphics[width=8cm]{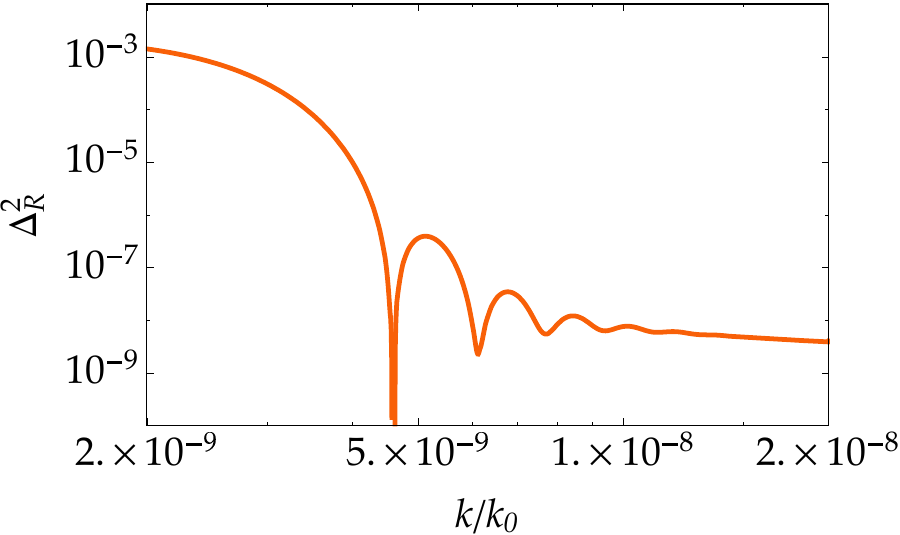} 
\includegraphics[width=8cm]{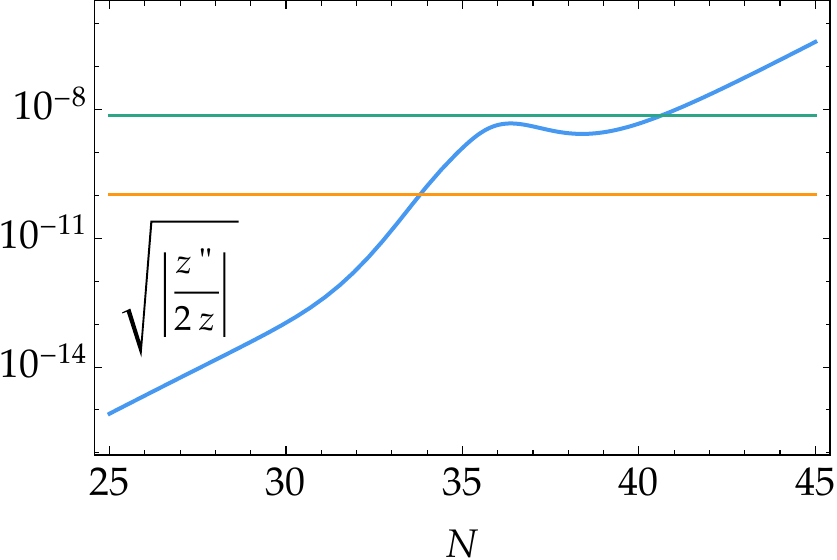} 
\includegraphics[width=8cm]{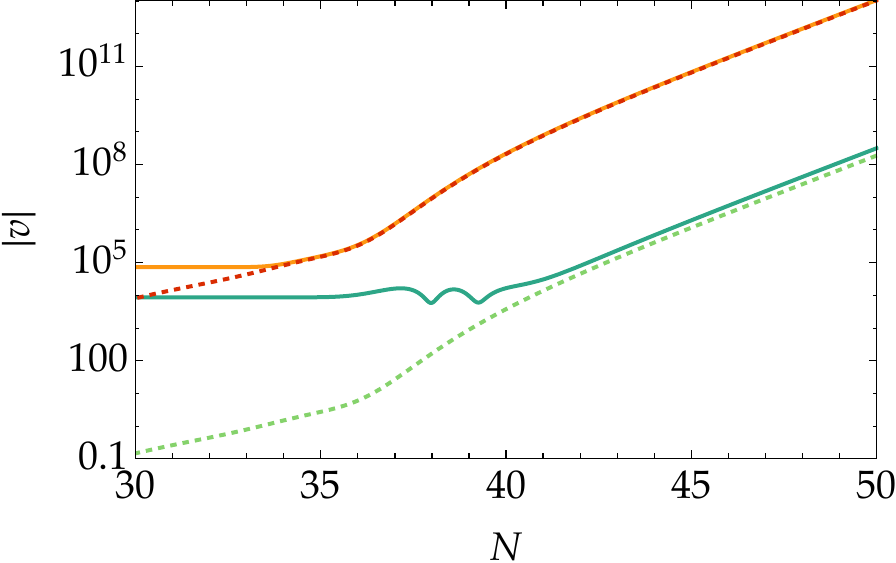} 
\includegraphics[width=8cm]{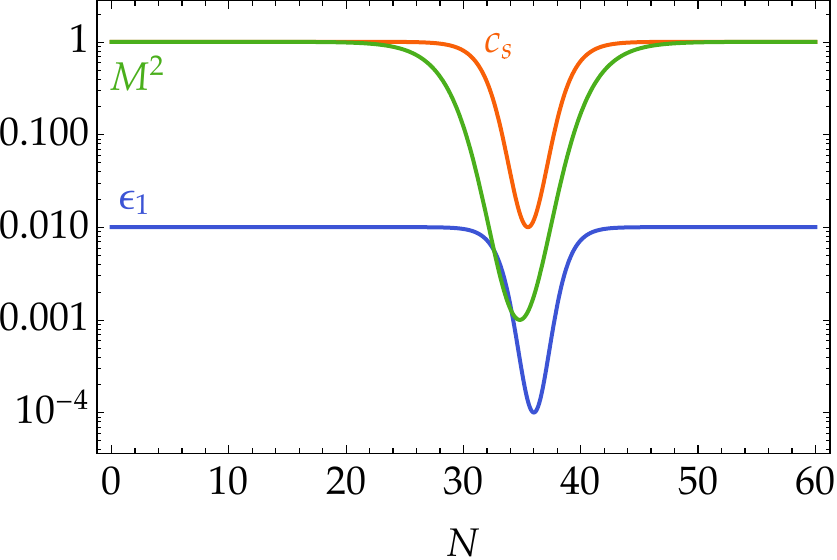} 
\includegraphics[width=8cm]{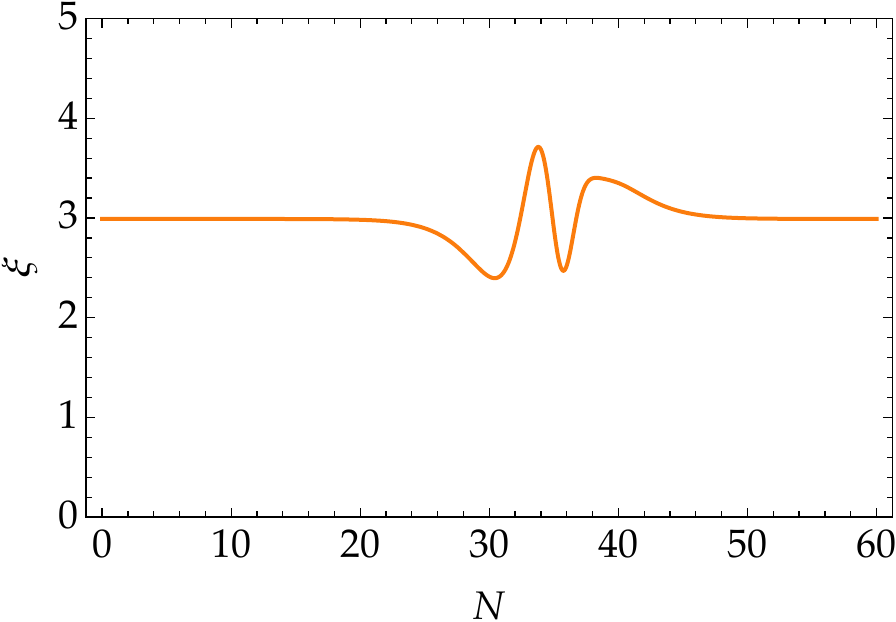}  
\caption{In the upper left panel we show the power spectrum $\Delta^2_{\mR}$ (with $k$ normalized to $k_0=H_{\rm end}$) computed from the exact numerical solutions (solid-red) and the analytical approximation (dashed-greed), ignoring multiple crossings, for the parametrization corresponding to $n_{\cs}=-2$, $n_\epsilon=-4.5$, $n_{M^2}=-3$, $\sigma_{\cs}=2.5$, $\sigma_\epsilon=2$, $\sigma_{M^2}=4$, $N_{\cs}=35.5$, $N_{\epsilon}=36$, $N_{M^2}=34.8$. The evolution $\epsilon$, $\cs^2$ and $M^2$ is shown in the bottom-left panel and the corresponding friction term is shown in the bottom-right panel, where we see that it remains positive so that the ``decaying'' mode is never activated and remains negligible. The middle-left panel shows the function determining the horizon crossing as explained in the main text; see \eq{eq:crossing}. We also display two values of $k$ whose solutions are shown in the middle-right panel, where we can see the exact numerical solutions (solid lines) and the approximate (super-horizon) solutions $\vert v\vert={z}/(\sqrt{k}\, {z_*})$ obtained by matching with the sub-horizon modes at the crossing time as explained in detail in the main text. We confirm that the matching gives a very good approximation to the super-horizon solution even if the slow-roll parameters change significantly at the crossing (red/orange curves). We also plot for illustrative purposes a mode that only presents one crossing, but passes very near the {\it horizon} before exiting. For this kind of mode, we also see oscillations expected for modes with multiple crossings (which means that they are not in the usual Bunch-Davis vacuum before definitely exiting the horizon). See the upper right panel. The approximate solution $\vert v\vert={z}/(\sqrt{k}\, {z_*})$ is less accurate for such modes.}
\label{Fig:M1C}
\end{figure}

\subsection{Phenomenological parametrization}\label{Sec:parametrization}

In order to illustrate the analytical results obtained the precedent sections, we will consider now a parameterization of $\cs$, $\epsilon$ and $M^2$ that allow for localized dips and obtain the power spectra in a couple of examples by numerically solving the Mukhanov-Sasaki equation. We will then compare the numerical solutions with the approximate analytical expressions found above and, in particular, we will highlight the importance of appropriately computing the crossing time for a correct matching. The parametrizations that we will use are constructed in terms of the function
\be
G(N;n_1,n_2,N_s,\sigma)\equiv(n_1-n_2)\tanh^2\left(\frac{N-N_s}{\sigma}\right)+n_2
\ee
{being} $N$ the number of e-folds. This function tends to $n_1$ sufficiently far from $N_s$, meaining $|N-N_s|\gg\sigma>0$, and becomes equal to $n_2$ around $N_s$ for a number of e-folds controlled by $\sigma$. Since we are interested in scenarios where deviations with respect to the standard case occur somewhere between the CMB scales and the end of inflation, we will typically have $N_s$ to be less than about 50-60 e-folds before the end of inflation. We parameterize $\epsilon$, $\cs$ and $M^2$ as
\begin{align}
 &\log_{10}\epsilon=G(N;n_\epsilon^{(0)},n_\epsilon,N_\epsilon,\sigma_\epsilon),\\ 
 &\log_{10}\cs=G(N;n_{\cs}^{(0)},n_{\cs},N_{\cs},\sigma_{\cs}),\\
 &\log_{10}\frac{M^2}{\mpl^2}=G(N;n_{M^2}^{(0)},n_{M^2},N_{M^2},\sigma_{M^2}).
 \end{align}
We should notice that the parameterizaton of $\epsilon$ also determines the background expansion through $H(N)=H_{\rm end}\exp[-\int_{60}^N \epsilon(N)\dd N]$, where $H_{\rm end}$ is the value of the Hubble parameter at $N=60$. 

In what follows we will choose $n_\epsilon^{(0)}\simeq -2$ (so that the deviation of the scalar spectral index from 1 is close to the one measured by CMB experiments at the corresponding scales), $n_{\cs}^{(0)}=0$ (i.e.\ $\cs\simeq 1$ sufficiently before the end of inflation, which corresponds to a canonical scalar field) and $n_{M^2}^{(0)}\simeq 1$ so that $M^2$ reduces to the usual Planck mass at sufficiently early times. At intermediate values of the number of e-folds --in practice, at some point between the CMB scales and the end of inflation-- $\epsilon\simeq 10^{-n_\epsilon}$, $\cs\simeq 10^{-n_{\cs}}$ and $M^2\simeq 10^{-n_{M^2}}\mpl^2$ near $N=N_\epsilon$, $N=N_{\cs}$ and $N=N_{M^2}$, respectively.
 
  \begin{figure}[t]
\includegraphics[width=8.25cm]{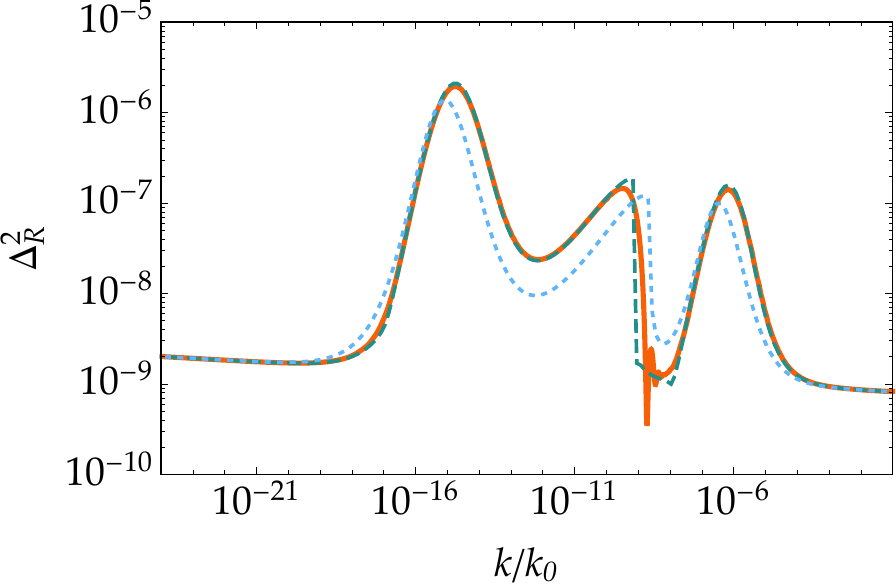} 
\includegraphics[width=7.75cm]{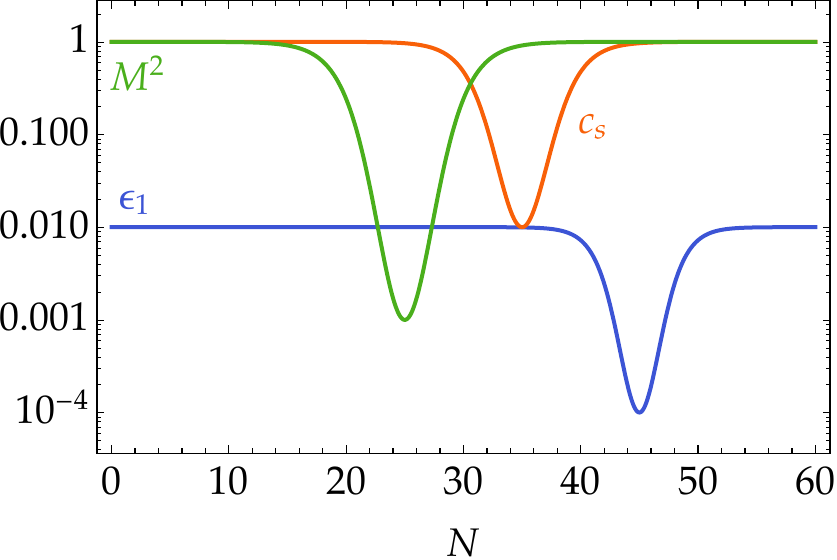} 
\caption{The left panel shows the power spectrum (with $k$ normalized to $k_0=H_{\rm end}$)   for the parameters displayed in the right panel that correspond to  $n_{\cs}=-2$, $n_\epsilon=-4$, $n_{M^2}=-3$, $\sigma_{\cs}=3.2$, $\sigma_\epsilon=2.5$, $\sigma_{M^2}=3.5$, $N_{\cs}=35$, $N_{\epsilon}=45$, $N_{M^2}=25$. The curves for the power spectrum in the left panel correspond to the exact numerical solution (red-solid), the analytical approximation with the crossing time computed from $k^2=a^2H^2/\cs^2$ (blue-dotted) and the analytical approximation using (\ref{eq:crossing}) to compute the crossing time (although neglecting multiple crossings). In the usual slow-roll regime, the three curves agree with high precision. As we approach the region where $\epsilon$, $\cs^2$ and $M^2$ change substantially, the inaccurate computation of the crossing times makes the blue-dotted curve deviate significantly from the exact solution, while an appropriate computation of the crossing gives by using (\ref{eq:crossing}) a much more accurate result. Since we neglect multiple crossings, the region where the oscillations arise from such crossings is not well-captured by the approximations.}
\label{Fig:M1Bi}
\end{figure}   

Having introduced the parameterization that we will use, we will now obtain the super-horizon primordial power spectrum at the end of inflation as a proof-of-concept, which will furthermore allow us to check the validity of some of our analytical expressions. We will focus on the case where the ``decaying'' mode never becomes a growing mode, i.e., $\xi$ does not turn negative for any $N$, and the amplification in the power spectrum is entirely due to the standard constant mode for $\mR$.  In the explicit single field inflationary model explored in Section \ref{sec:single_field_models} we will see an example where the decaying mode turns intro a growing one. Here, we will discuss two particular examples leading to a {\it cooperative} and a {\it multi-modal} enhancement of the power spectrum. 

In Figure  \ref{Fig:M1C}  we summarize the first explored case where $\epsilon$, $\cs^2$ and $M^2$ feature an approximately simultaneous dip, cooperating to produce a larger enhancement in the power spectrum than what they would produce individually. We see that far from the location ($N\simeq 35$) in the feature from $\epsilon_1$, $\cs$ and $M^2$ we recover the usual nearly flat power spectrum. On the other hand, near the location of the feature the power spectrum is enhanced, giving rise to a peak whose value is in agreement with the expectation from (\ref{eq:PSsr}). Finally, we can observe how a given band of modes will experience multiple crossings, what will result in the oscillations that we can see toward the end of the peak in the power spectrum, again in accordance with our expectation from (\ref{Eq:Pkosc}). Let us recall that the number of e-folds of inflation that can be probed with the CMB is $\sim 10$ (and a few more with LSS  and spectral distortions), so that such a feature in $\Delta_\mR^2$ would be impossible to probe with such observations. The generation of PBHs at such small scales could then be a possibility to test the physics of inflation for (distance) scales that re-enter the horizon much earlier, as in this example.

In Figure \ref{Fig:M1Bi} we show a different example, for which the dips in $\epsilon$, $\cs$ and $M^2$ are sufficiently apart as to give rise to three distinct peaks in the power spectrum. This example also has a band of modes that undergo several crossings, leading to a characteristic pattern of oscillations in the power spectrum in between the last two peaks. The friction parameter again remains positive, so that the analytical expressions obtained above based on the constant mode are again valid. In this case, however, some slow parameters present large variations that make the function $\Xi$ be relevant for the computation of the crossing time. We have plotted the power spectra obtained from the exact numerical solution and two different analytical expressions. The blue-dashed line corresponds to (\ref{eq:PSsr}) neglecting $\Xi$, i.e., the crossing time is computed from $k=aH/\cs$, whereas the green-dotted curve is computed by taking into account $\Xi$. We can clearly see the importance of appropriately computing the matching time, in which case the analytical expression gives an accurate result for the power spectrum, even when the quantities $\epsilon$, $\cs$ and $M^2$ present large variations. Since we have ignored the effects from having multiple crossings, we see that the analytical expression gives a much worse result for the region with oscillations.

\section{The EFT of inflation and PBH formation} \label{EFTPBH}

The EFT of inflation \cite{Cheung:2007st} is a general description of the dynamics of small fluctuations around a time-dependent quasi-de Sitter background. The action for these fluctuations can be constructed in the so-called unitary gauge, by writing --in a hierarchical way determined by their number of derivatives-- the independent combinations of the spacetime metric that are invariant under spatial diffeomorphisms. At lowest order in derivatives, the quadratic action for scalar fluctuations is precisely \eq{Eq:quadraticgeneral}, which describes all the models whose Lagrangian is of the form $\mathcal{L}=p(\phi,X)$, where $\phi$ is a real scalar, 
$X \equiv - \partial_\mu \phi\, \partial^\mu \phi /2$ and $p$ is an arbitrary function (but able to sustain inflation). Higher order derivative terms contributing to the effective  action for fluctuations are also allowed, but are supposed to be subdominant, within the logic of the EFT. 

It is common to recast the action for scalar fluctuations in terms of a field $\pi=-H\,\zeta$, which is the Goldstone boson arising from the breaking of time diffeomorphisms, due to inflation. In terms of this variable, the first cubic interaction that can be written is of the form $(\partial_i\pi)^2 \dot\pi/a^2$ and it comes with a coefficient that (due to the symmetries) is entirely fixed by the quadratic action, and can be written as a function of $\epsilon$, $\cs$ and $H$. This interaction leads to the following estimate of the (partial wave) unitarity cut-off \cite{Cheung:2007st}, denoted by $\Lambda$:
\begin{align}
\Lambda^4\sim 16\pi^2 M_P^2 H^2 \epsilon \frac{\cs^5}{1-\cs^2}\,.
\end{align}
The only other possible cubic interaction that can appear in the action for $\pi$ is just $\dot\pi^3$ but, unlike for the previous one, its coefficient is not completely determined by the symmetries. Without any loss of generality, we can write the ratio of the two cubic terms as ${\alpha}{(\partial_i\pi)^2}/{(\cs^2 a^2\dot\pi^2)}$,
where $\alpha$ is a dimensionless function of time. Then, the fourth power of the unitarity cutoff due to $\dot\pi^3$ is $\sim  \alpha^2 \Lambda^4$  \cite{Senatore:2009gt}, which can be larger than $\Lambda^4$ provided that $\alpha \gtrsim 1$. 

Since the interaction  $(\partial_i\pi)^2 \dot\pi/a^2$ is always present in the EFT for $\cs\neq 0$; if we assume slow-roll and impose $\Lambda \gg H$ we get an inequality between the speed of sound and the amplitude of the primordial spectrum of $\zeta$:
\begin{align} \label{eftbound}
\cs^4\gg \Delta_\zeta^2\,,
\end{align}
so that $\cs$ cannot be smaller than $\sim 0.3$ for $\Delta_\zeta^2\sim 10^{-2}$. Taken at face value, and considering standard values for $\epsilon\sim\mathcal{O}(1-n_s)\sim\mathcal{O}(0.01)$, this means that in the context of the {(slow-roll)} EFT of inflation  the production of a sizable population of PBHs that could be relevant for the DM problem cannot be achieved from slow-roll inflation through a very small speed sound.  

There are a couple of factors that one could think might help to alleviate or even circumvent this obstruction. First of all, the condition $\Delta_\zeta^2\sim 10^{-2}$ for having an $\mathcal{O}(0.1)$ contribution to $\Omega_{\rm DM}$ is not, strictly speaking, a bulletproof prediction. This condition arises from the application of the general formalism for PBH formation from the collapse of large (and rare) radiation density fluctuations, as described in Section \ref{largecollapse}. However, there exist reasonable arguments suggesting that this framework for PBH formation may need to be modified --see e.g.\ \cite{Yoo:2018kvb,Germani:2018jgr} for recent works in this direction-- and a smaller value of $\Delta_\zeta^2$ could be sufficient to obtain a large DM abundance. In particular, it is important to remark that one of the main assumptions of the formalism of Section \ref{largecollapse} is that the primordial spectrum is Gaussian; see equation \eq{GPS}. However, a small $\cs$ generically induces large non-Gaussianities, simply because the interactions in the EFT of inflation are controlled by positive powers of $1/\cs^2$. Depending on the shape of these non-Gaussianities the threshold for collapse could effectively be lowered, thus favoring the formation of PBHs and hence leading to a higher abundance than in the Gaussian estimate. It has been argued in \cite{Franciolini:2018vbk} that local non-Gaussianities arising in an ultra slow-roll regime \cite{Kinney:2005vj} with $\cs=1$ can make PBH production far more inefficient than what the Gaussian assumption predicts. Still, other non-Gaussian shapes could go in the opposite direction. A general study of the effect of non-Gaussianities on PBH formation with a varying $\cs\neq 1$ would be required to clarify this issue, but it goes beyond the scope of our present work. 

In spite of the previous considerations, it has to be noted that even if a much lower primordial amplitude such as e.g.\ $\Delta_\zeta^2\sim 10^{-5}$ were enough for $\Omega_{\rm PBH}\sim 0.1$ --which seems rather far-fetched, given our current theoretical knowledge-- the bound \eq{eftbound} would nonetheless imply $\cs\gtrsim0.05$. This is still  a large lower bound for the purpose of PBH formation from a small $\cs$, since it requires $\epsilon\sim\mathcal{O}(10^{-5})\ll \cs$. Indeed, if we assume that generalized slow-roll holds --both for the CMB scales and for those relevant for PBH formation-- the bound \eq{eftbound} implies
\begin{align} \label{cmbb}
\epsilon_{\rm  PBH}\ll {\Delta_{\zeta\,{\rm CMB}}^2}\, c_{s\, {\rm CMB}}  \left(\frac{r_{\rm CMB}}{0.07}\right)\left(\frac{\Delta_{\zeta\,{\rm PBH}}^2}{0.01}\right)^{-5/4}\,,
\end{align}
which means that for $r_{\rm CMB}\sim 0.07$ --which is the current upper limit \cite{Akrami:2018odb}-- and $c_{s\, {\rm CMB}}\sim 1$  it needs to be mostly $\epsilon$ (and not $\cs$) that leads to the PBH generating peak in the primordial spectrum, even if the required $\Delta^2_{\zeta\,{\rm PBH}}$ were as low as $\sim 10^{-7}$. The idea of PBH production from a very small $\epsilon$ appears then to be natural from the point of view of the EFT, whereas achieving the same with small $\cs$ is, at first sight, incompatible with the EFT. 

The idea of PBH formation with a small $\cs$ can nonetheless be explored further, still in the EFT framework. A possibility to avoid the bound \eq{eftbound} within the EFT of inflation is to consider models with a dispersion relation $\omega \sim k^{2n}$ where $n$ is some positive number. The best known example of this (with $n=1$) is ghost inflation \cite{ArkaniHamed:2003uy,ArkaniHamed:2003uz}. This corresponds to a generalized $p(\phi,X)$ model in which the scalar field has constant velocity, breaking Lorentz symmetry ``spontaneously''. This can lead to inflation provided that the derivative of $p$ with respect to $X$ tends to zero dynamically. In this regime, the coefficient of the term $(\partial_i\pi)^2$ in the effective action becomes zero, which corresponds to $\cs\rightarrow 0$. The quadratic action can then be stabilized provided that higher order derivative terms come into play in the Lagrangian for $\phi$, leading to a quadratic term for $\pi$ that goes as $(\partial^2\pi)^2$, which comes associated to a new mass scale, $M_{\rm GI}$, so that the dispersion relation is $\omega\sim k^2/M_{\rm GI}$. A proper power counting shows that this scale $M_{\rm GI}$ is the ultraviolet cut-off of the model in this regime. Interestingly, the primordial spectrum scales as
\begin{align} \label{GIPS}
\Delta_\zeta^2\sim 0.02\left(\frac{H}{M_{\rm GI}}\right)^{5/2}\,,
\end{align} 
whereas the level of (equilateral) non-Gaussianities is $f_{NL}\lesssim 100$, where we are assuming that precisely the same mass scale that appears with $(\partial^2\pi)^2$ controls also the relevance of $\dot\pi(\partial_i\pi)^2$, see \cite{ArkaniHamed:2003uz} for details. The estimate \eq{GIPS} tells us that ghost inflation may be able to produce $\Delta_\zeta^2$ one or two orders of magnitude below $10^{-2}$, given that the consistency of the EFT always requires $H\ll M_{\rm GI}$. Since the level of non-Gaussianities is substantial, it may be that such a primordial amplitude could be sufficient to generate $\Omega_{\rm PBH}\sim 0.1$, although a detailed analysis needs to be done. Therefore we identify ghost inflation as a potentially interesting possibility for the generation of abundant PBH DM. Notice that in this context we only need ghost inflation to operate during PBH formation and the generation of the CMB scales could occur within standard slow-roll, with a transition between the two regimes at intermediate scales. It is interesting that if we assume that $H$ does not change significantly during inflation and its value is such that it saturates the current bound on the tensor-to-scalar ratio, we get that $M_{\rm GI}$ { is well above the GUT scale:
\begin{align}
M_{\rm GI}\gg H\sim 5\times 10^{13}\, {\rm GeV}\,.
\end{align}
}
Although ghost inflation dynamics appears as a promising possibility for PBH formation within the EFT of inflation, eventual models with dispersion relations of higher order ($w^2\sim k^{2n}$, $n\geq3$) happen to suffer from an even worse problem as the vanilla EFT of inflation. This can be seen considering the scaling dimensions of $\pi$, which in such a case implies that the operator $\dot\pi(\partial\pi)^2$ is strongly coupled in the infrared (i.e.\ for $k\ll aH$) \cite{Cheung:2007st}.

A more drastic possibility for evading the obstruction on PBH formation from a small $\cs$ imposed by \eq{eftbound} is to focus on models of inflation that can be considered ultraviolet complete, {outside of the realm of the EFT of inflation.} This is what we do in the next section. {In Section \ref{pbhsolid} we will consider PBH formation in the context of solid inflation, which is also outside the EFT of inflation.}

\section{\label{sec:single_field_models} A single-field toy model}

As we have already mentioned earlier, speeds of sound different from that of light are generic in models with non-standard kinetic terms \cite{ArmendarizPicon:1999rj,Garriga:1999vw}. Let us consider a Lagrangian of the form $\mathcal{L} = p(X,\phi)$ for a real scalar $\phi$, with $X \equiv - \partial_\mu \phi\, \partial^\mu \phi /2$. The two independent Einstein equations for a solution $\phi=\phi(t)$ are 
\begin{equation}
	\label{eq:Friedman}
	3 M_P^2 H^2  = \rho = 2Xp_{X} - p \, , \qquad -2 M_P^2 \dot{H} = p + \rho = 2 X p_{X}\,,
\end{equation}
where the subscripts indicate partial derivatives. The time functions $\epsilon$ and $\cs^2$ are 
\begin{equation}
	\label{eq:epsilon_def}
	\epsilon= -\frac{\dot{H}}{H^2} = \frac{X p_{X}}{M_P^2H^2}\,,\quad \cs^2 = \frac{p_{X}}{\rho_{X}}= \frac{p_{X}}{p_{X} + 2X p_{	XX}} \,,
\end{equation}
which we will assume to be non-negative. In the context of inflation, a well-known example of this kind of models is DBI inflation~\cite{Silverstein:2003hf,Alishahiha:2004eh} which can be motivated in string theory considering a D3-brane that can be effectively described~\cite{Leigh:1989jq} by a DBI action~\cite{Dirac:1962iy,Born:1934gh}. The possibility of deriving this type of models from string theory compactifications argues in favor of their radiative stability \cite{Tseytlin:1999dj,Silverstein:2003hf}, distinguishing them from arbitrary $p(\phi,X)$ Lagrangians.  A generalization of these models is given by
\begin{equation}
	\label{eq:lagrangian_for_dbilike}
	p(\phi,X) = -V+	\left(1 - \sqrt{1 - 2 f\, X}\right)f^{-1}  \, ,
\end{equation}
where the potential $V$ and the warp factor $1/f$ are so far unspecified functions of $\phi$. In this model 
\begin{equation}
	\cs^2 = 1 - f\, \dot{\phi}^2  \,,\quad \epsilon = \frac{3}{2} \, \left[ \frac{1-\cs^2 }{1 + \left( f\, V -1 \right) \cs} \right] \, . 
\end{equation}
For $f\,\dot{\phi}^2 \simeq 1$ with $f \gg M_P^{-4}$ it is possible that $\cs \ll 1$ and $\epsilon \ll 1$ simultaneously, achieving dynamically a strong enhancement of the scalar primordial power spectrum. Given a potential $V$ which is able to sustain inflation, this can be realized if $f$ has a localized peak at some point, say $\phi_0$, corresponding to a scale between the CMB and the minimum of $V$ where reheating takes place. Around such a point $\dot{\phi} \ll 1$, so that the field remains ``stuck'' at $\phi_0$ for several e-folds.\footnote{This mechanism can reduce $n_s$ and enhance $r$ with respect to the case $f(\phi)M_P^{4}=1$; see ~\cite{Domcke:2016bkh}.}

 \begin{figure}[t]
\includegraphics[width=8.1cm]{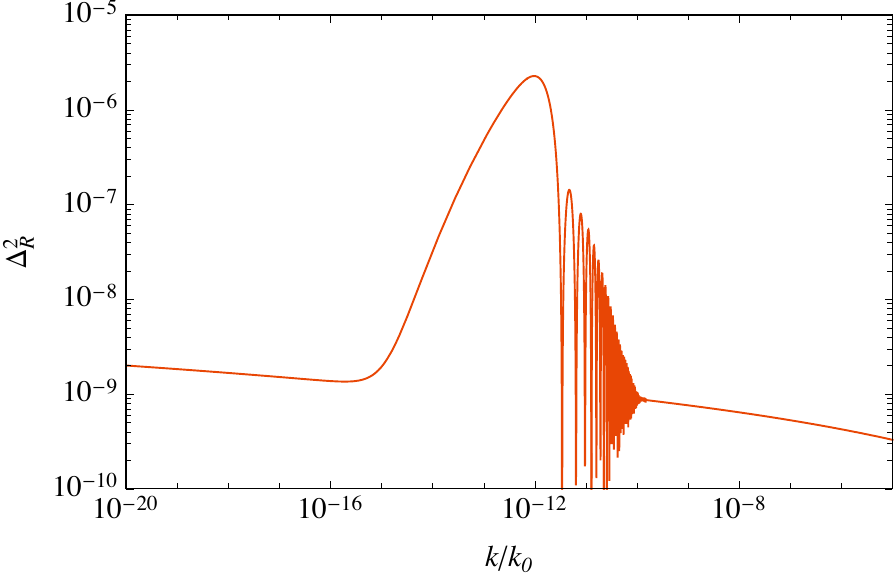}\hspace{0.1cm}
\includegraphics[width=8cm]{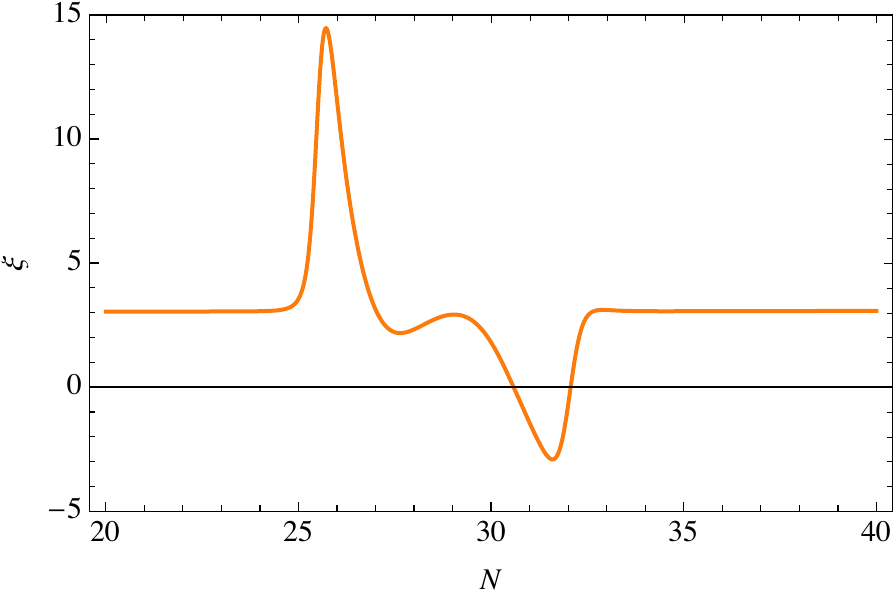}
\includegraphics[width=8cm]{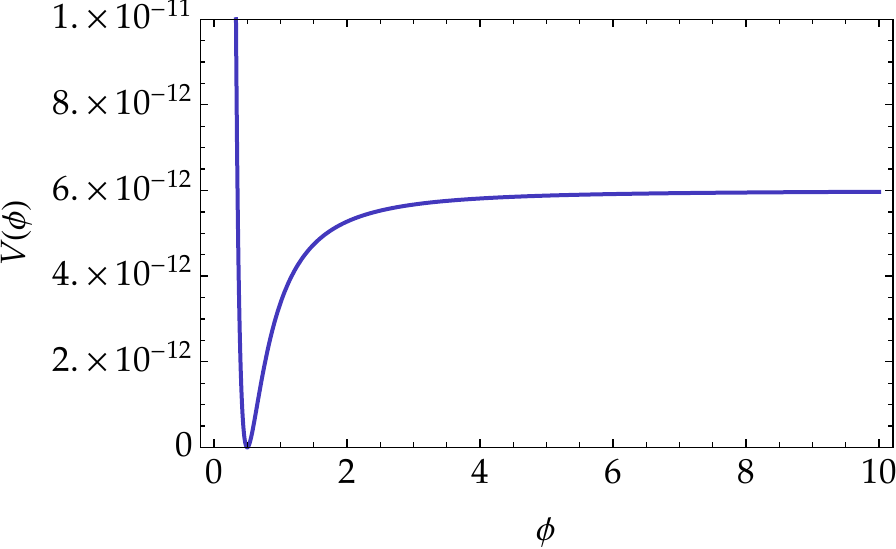} \hspace{0.1cm}
\includegraphics[width=8cm]{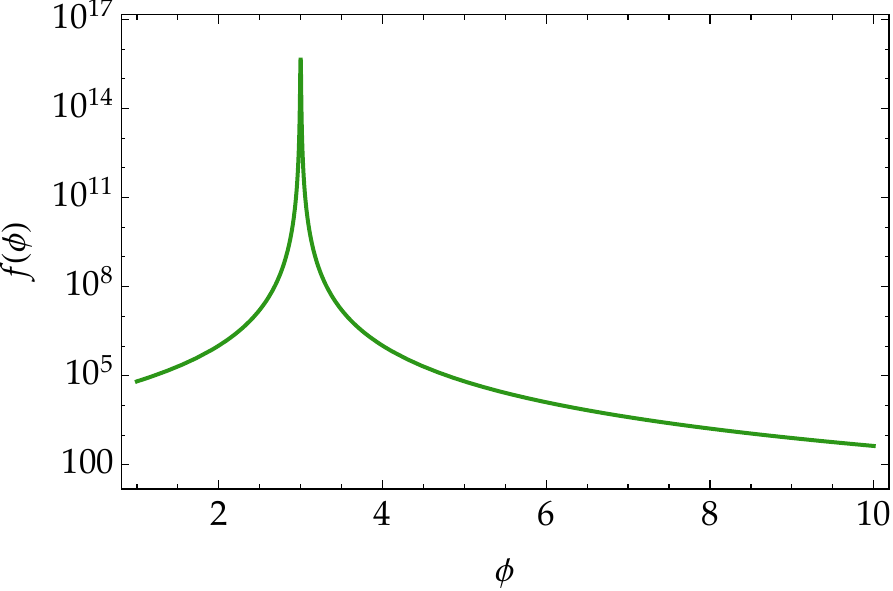} 
\caption{Upper left panel: Primordial power spectrum $\Delta_\mR^2$ for the single field model described by the equations \eq{eq:lagrangian_for_dbilike} and \eq{example1} for $V_0 = 9.6 \times 10^{-11}M_P^4 $, $y = 0.25$ --so that for $f(\phi) = M_P^{-4}$, $n_s \simeq 0.974$ and $r \simeq 0.002$--, $q = 2 \times 10^{-16}$ and $Q = 10^{-6}$. The value of $V_0$ is chosen to respect the COBE normalization and $k_0$ is the value of $H$ at the end of inflation. Upper right panel: friction term, defined in \eq{frictious}, which becomes negative for some e-folds, activating the ``decaying'' mode discussed in Section \ref{Sec:Decaying}. Lower panels: The corresponding potential $V(\phi)$ and the function $f(\phi$). The potential produces the usual slow-roll phase at large values of the field. Inflation ends and is followed by oscillations of $\phi$ near the minimum that will give rise to reheating. The ``spike'' on $f(\phi)$ around $\phi\simeq 3 M_P$ is the feature responsible for PBH formation.}
\label{Fig:example}
\end{figure}

To illustrate this mechanism, let us consider:
\begin{equation} \label{example1}
	V(\phi) = V_0 \left(1 - y\frac{M_P^2}{\phi^2} \right)^2 \, , \qquad \qquad \frac{1}{f(\phi)} = {q\,M_P^4 + Q \left(\phi - \phi_0\right)^4}  \ ,
\end{equation}
where $V_0$, $y$, $Q$ and $q$ are positive constants and $\phi_0$ is the value of $\phi$ at which the maximum of $f$ is localized. This is a rather minimal modification of the usual expression in DBI inflation: $f \propto \phi^{-4}$. The choice of $V$ is not crucial for the mechanism. The one above has been taken for simplicity, although it gives for $f=1$ a slightly bluer spectrum than that measured by Planck \cite{Akrami:2018odb}.

The level of non-Gaussianities in $p(\phi,X)$ models can be characterized with the aid of the following two quantitites   \cite{Chen:2006nt,Chen:2010xka}: $ \Sigma =X p_{X} + 2 X^2 p_{XX}$ and $\lambda =  X^2 p_{XX} + ({2}/{3})  X^3 p_{XXX}$. In general models of DBI inflation the leading non-Gaussian contributions grow as $1/\cs^2$ and $\lambda/\Sigma$, which in our specific example is $\lambda/\Sigma=f\,X/\cs^2$. The effect of non-Gaussianities should be included in an estimate of the PBH abundance produced with this mechanism, although we focus here only on the (linear) enhancement of the primordial spectrum.

In Figure \ref{Fig:example} we show the primordial power spectrum and the functions $f$ and $V$ for a concrete choice of parameters. As expected, as soon as the field approaches $\phi_0$ (which here is equal to $3 M_P$), the spectrum is strongly amplified due to the temporal decrease of $\epsilon$ and $\cs$, which take place simultaneously, giving an explicit realisation of a cooperative enhancement. Moreover, the non-trivial evolution of $\cs$ leads to a band of $k$-modes undergoing multiple crossings, which results in the oscillations that we can see in the power spectrum.  At scales which are sufficiently apart from the peak, the spectrum shows the usual nearly scale invariant form. While the amplitude of the peak shown in this figure is not sufficient to trigger the generation of a significant PBH population, in general it is possible to choose the parameters of the model in order to get larger values. In this work we do not study in detail the generated PBH population and the example presented in this section should be intended as a proof-of-concept. A dedicated analysis of the parameter space and of the associated PBH populations is left for future works. Lager values of the amplitude of the peak are obtained by imposing a sharper variation of $f$ around $\phi_0$. In such a case, more modes will experience multiple horizon crossings and, consequently, more oscillations will arise in power spectrum. Resolving with sufficient accuracy all the oscillations becomes very costly (from a numerical point of view) as the number of oscillations increases.

\section{PBH formation in solid inflation} \label{pbhsolid} 
Solid inflation~\cite{Gruzinov:2004ty,Endlich:2012pz} is a model of inflation that features a trio of scalar fields $\Phi^i(t,x^j)$ that are derivatively coupled and respect an internal $SO(3)$ symmetry. When coupled to gravity in a minimal way, this system becomes the relativistic  EFT description of an elastic solid. The fields $\Phi^i$ are the only degrees of freedom needed to describe the solid at low energies and represent its Lagrangian (or comoving) coordinates. They can also be viewed as the St\"uckelberg fields of broken spatial diffeomorphisms, since in their background configuration $\Phi^i=x^i$. Interestingly, this static background is compatible with a (time-dependent) FLRW metric and, in particular, it may sustain a quasi-de Sitter phase. At lowest order in the derivative expansion that defines the EFT, the dynamics of the solid is governed by the following action: 
\begin{align} \label{acts}
S=\frac{M_P^2}{2}\int\,  \textrm{d}^4x \,\sqrt{-g}\, R + \int\,  \textrm{d}^4x\, \sqrt{-g}\, F(X,Y,Z)\;,
\end{align}
where $F$ is some function of three independent operators: $X$, $Y$ and $Z$. In this section we follow the notation of~\cite{Endlich:2012pz} for these operators.\footnote{Notice that in previous sections we used the symbol $X$ for the kinetic term of a single scalar.} These are chosen for convenience as the following functions of the Cayley invariants  of the matrix $B^{ij}=\partial_\mu\Phi^i\partial^\mu\Phi^j$:  $X=\text{Tr} B$, $Y=\text{Tr}(B^2)/X^2$ and  $Z=\text{Tr}(B^3)/X^3$. This choice of operators allows to write the energy density and the isotropic pressure of the solid only in terms of $F$ and its variation with respect to $X$. Concretely: $\rho = -F$ and $p=F-(2/3)XF_X$. Therefore, the equation of state of the fluid, $w$, and the slow-roll parameter $\epsilon$ are related through $F$ and $F_X$ evaluated on the background:
\begin{align}
\epsilon=\frac{3}{2}(1+w)=\frac{X F_X}{F}\;.
\end{align}
Inflation takes place if the relative variation of $F$ with respect to $X$ is small enough that $\epsilon<1$. 

There are some peculiar aspects of inflation driven by the action~\eqref{acts} that make it potentially interesting in the context of PBH formation. First of all, the gauge invariant quantitites $\zeta$ and $\mathcal{R}$ evolve in time for $k\ll a\, H$ (unlike for standard inflation models) and are different from each other in this limit. Since PBHs form when sufficiently large fluctuations re-enter the ``horizon'' ($k = a\, H$) after inflation has ended, this super-horizon evolution has to be taken into account. The dynamics of $\zeta$ and $\mathcal{R}$ obeys the following system of equations at linear order in perturbations: 
\begin{align} \label{E1z} 
\dot{\mathcal{R}}+(3+\epsilon_2-2\epsilon)H\mathcal{R}+3 H c_L^2 \zeta =0\,,\quad
\dot \zeta+\epsilon H\zeta-\epsilon H \left(1+\frac{k^2}{3a^2H^2\epsilon}\right)\mathcal{R}=0\,.
\end{align}
In these expressions $c_L^2$ denotes the speed of propagation of longitudinal phonons in the solid  and whose precise dependence on $F$ and its derivatives will be discussed later. For the time being, it is enough to point out that although the background evolution is insensitive to $Y$ and $Z$, the fluctuations around this background are not, due to $c_L^2$. The equations~\eqref{E1z} are valid even if no slow-roll conditions are assumed. Combining them we can obtain a second order differential equation for each variable separately:
\begin{align} \label{eqR}
 \ddot{\mathcal{R}}+ &  (3-2s_L+\epsilon_2)H\dot{\mathcal{R}}  \nonumber \\
&+\left[\frac{c_L^2}{a^2}k^2+2\left(2\epsilon-\epsilon_2-3\right)s_L H^2 +\left(3+3c_L^2-2\epsilon-\epsilon_2\right)\epsilon H^2+\epsilon_2\epsilon_3 H^2\right]\mathcal{R}=0\;.  
\end{align}
and
\begin{align}\label{eqz} 
\ddot\zeta + & \frac{(5+\epsilon_2-2\epsilon)k^2+9a^2H^2\epsilon}{k^2+3a^2H^2\epsilon}\,H\dot\zeta  \nonumber \\
&+ \left[\left(\frac{c_L^2}{a^2}+\frac{5-4\epsilon+2\epsilon_2}{k^2+3a^2H^2\epsilon}H^2\epsilon \right)k^2+ 3H^2\epsilon\left(c_L^2+\frac{3-2\epsilon+\epsilon_2}{k^2+3a^2H^2\epsilon}a^2H^2\epsilon\right)\right]\zeta=0\;,
\end{align}
where 
\begin{align}
s_L=\frac{\dot c_L}{c_L\, H}\,.
\end{align} 
In the limit $k\rightarrow 0$ both equations feature a non-derivative term that makes $\mathcal{R}$ and $\zeta$ evolve outside the horizon. Such a term is absent in the standard equations for a single scalar  with a speed of sound $\cs \neq 1 $. In that case the evolution of $\mathcal{R}$ is given by: $\ddot{\mathcal{R}}+\left(3-2\,s+\epsilon_2\right)H\dot{\mathcal{R}}+{\cs^2}k^2 \mathcal{R}/{a^2}=0$, which is equation \eq{massiveR} for $m=0$ and $M=M_P$. Clearly, the non-derivative term of this equation vanishes in the limit $k\rightarrow 0$. As discussed in ~\cite{Endlich:2012pz}, the origin of the super-horizon behavior of the fluctuations in solid inflation can be traced back to the existence of a small anisotropic stress in the system.\footnote{ It is worth noting that the model of gaugid inflation \cite{Piazza:2017bsd} (based on three $U(1)$ gauge vector fields with an internal $SO(3)$ global symmetry) has many similarities with solid inflation and much of our conclusions can be translated there. In particular, inflation is supported by a inhomogeneous background configuration of the fields and, the primordial fluctuations also evolve in time for $k\rightarrow 0$ and the action for $\zeta$ has the same form in both models. }

The complications of describing PBH formation in solid inflation due to the super-horizon evolution of $\mathcal{R}$ and $\zeta$ can be partially simplified by assuming that the system satisfies generalized slow-roll conditions. In this regime and for $k\ll a\, H$~\cite{Endlich:2012pz}:
\begin{align} \label{srs}
\mathcal{R}\simeq-\frac{H}{\sqrt{4  \epsilon c_L k^3}M_P}\,,\quad
\zeta\simeq \frac{H}{\sqrt{4  \epsilon c_L^5 k^3}M_P}\;,
\end{align}
where we are neglecting, subdominant, time-dependent contributions that are  slow-roll suppressed. Leaving aside these time dependencies, the expression for $\mathcal{R}^2$ is the same as that in ``standard'' single-field inflation with the replacement $\cs\rightarrow c_L\neq 1$. However, quite remarkably, $\zeta^2$ depends on the fifth power of $c_L$. Thus, moderately small values of $c_L$ allow a significant enhancement of $\zeta$ with respect to $\mathcal{R}$. Assuming that the slow-roll corrections  to~\eqref{srs} are indeed negligible, the corresponding primordial spectra at horizon exit can give a good estimate of the spectra when inflation ends. The dimensionless spectrum for $\mathcal{R}$ is defined as 
\begin{align}
\Delta_\mathcal{R}^2(k)=\frac{k^3}{2\pi^2}P_{\mathcal{R}}(k)=\frac{ \textrm{d} \sigma^2_\mathcal{R}}{ \textrm{d} \log k}\,.	
\end{align}
where the ensemble average $\langle|\mathcal{R}(\mathbf{k})\mathcal{R}(\mathbf{k'})|\rangle=(2\pi)^3\delta^{(3)}(\mathbf{k}+\mathbf{k'})P_{\mathcal{R}}(k)$, with $k=|\mathbf{k}|$ (and analogously for $\zeta$). Therefore, with the aforementioned approximation:
\begin{align} \label{appsrg}
\Delta_\mathcal{R}^2\simeq\frac{H^2}{8\pi^2  \epsilon c_L M_P^2}\simeq c_L^4\,\Delta_\zeta^2\;.
\end{align}
This result, together with equation \eq{xacte}, implies that the evolution of $\zeta$ and $\mathcal{R}$ after inflation  can have an important impact on the predictions of PBH formation in solid inflation. Indeed, the way in which inflation ends and the subsequent reheating process determine how the primordial fluctuations of $\zeta$ and $\mathcal{R}$ get converted into radiation fluctuations (and eventually to PBHs).

As we have noted, an interesting property of~\eqref{appsrg} for PBH formation (in comparison to the standard case of a single field) is the high power of $c_L$ in the denominator of $\Delta_\zeta^2$. This suggests that a not to small $c_L$ may be sufficient to generate a large PBH abundance, provided that the radiation density fluctuation inherits the enhancement of $\Delta_\zeta^2$ for $c_L\rightarrow 0$. Before discussing how this may occur, let us first see how a small $c_L$ can arise in solid inflation. 

The longitudinal sound speed can be written as
\begin{align}
c_L^2=c_a^2+\frac{4}{3}c_T^2\;,
\end{align}
where 
\begin{equation}
\begin{aligned}
c_T^2 =1+\frac{2}{3}\frac{F_Y+F_Z}{XF_X}\,,\quad
c_a^2 =-\frac{1}{3}+\frac{2}{3}\frac{X F_{XX}}{F_X}=-1+\frac{2}{3}\epsilon-\frac{1}{3}{\epsilon_2}
\end{aligned}
\end{equation}
are, respectively, the speed of transverse phonons and the ``adiabatic'' sound speed  $c_a^2= \dot p/\dot\rho$. During slow-roll inflation, $|\epsilon_2|\ll 1$ and $\epsilon\ll 1$ so that $c_a^2\simeq -1$\;. Then, in order to avoid $c_L^2\simeq -1$ we need $c_T^2\neq 0$ and sufficiently large (but subluminal). In general, there is sufficient freedom to obtain a small (and positive) $c_L^2$ by exploiting the dependence of $F$ on $Y$ and $Z$ to ensure that $4c_T^2\simeq -3c_a^2$. For instance, as we have just seen, if $\epsilon$ and $\epsilon_2$ are negligible, $c_L^2\simeq 0$ provided that $c_T^2\simeq 3/4$. In this case $\epsilon,|\epsilon_2|\ll c_L^2 \ll 1$. The converse hierarchy: $ c_L^2 \ll \epsilon,|\epsilon_2|\ll 1$ can be achieved in various ways. One of them corresponds to the solid entering temporarily a phase in which it becomes approximately invariant under diffeomorphisms of the $\Phi^i$ that preserve the internal volume, behaving as a quasi-perfect fluid, so that, in practice, $c_T\simeq 0$. Then, we would also have to require $2\epsilon-\epsilon_2\simeq 3$, which implies (at least) a mild violation of slow-roll. Interestingly, if inflation does not end during that period, $\epsilon\ll |\epsilon_2|$, and the system actually undergoes a phase of approximate ultra-slow-roll~\cite{Tsamis:2003px}, characterized by $-3\lesssim \epsilon_2$. This is similar to the way in which PBHs are formed in canonical single-field models which feature a very flat plateau; see e.g.\ \cite{Ballesteros:2017fsr}. In this situation, the slow-roll expressions~\eqref{srs} cannot be used reliably in general, although we still expect a significant enhancement of the fluctuations. In fact, for the single field case, it was found in~\cite{Ballesteros:2017fsr}  that the enhancement can be several orders of magnitude larger than what the slow-roll analysis would indicate. 

In the model of solid inflation as proposed in \cite{Endlich:2012pz}, inflation ends through a transition from a elastic solid to a perfect fluid. As we just mentioned, this is	 a phase that is invariant under internal diffeomorphisms that maintain the volume: $\Phi^i\rightarrow \hat\Phi^i(\Phi^j)$ with $\det[\partial\hat\Phi^i/\partial\Phi^j]=1$  and, consequently, the function $F$ of~\eqref{acts} depends only on the determinant $\det B=(1-3Y+2Z)X^3/6$. In this regime, the longitudinal speed of sound is $c_L^2= b F_{bb}/F_b$, where $b=\sqrt{\det B}$ and~\cite{Ballesteros:2012kv}:
\begin{equation}
\begin{aligned} \label{fzeta}
\ddot\zeta+\left(3-\frac{1+3 c_L^2}{1+3\epsilon a^2H^2/k^2}\right)H\dot\zeta+ 
 \left(k^2\frac{c_L^2}{a^2}-\frac{(1+3 c_L^2)\epsilon H^2}{1+3\epsilon a^2H^2/k^2}\right)\zeta=0\,.
\end{aligned}
\end{equation}
From this equation, we see that $\zeta$ becomes constant, as expected, in the super-horizon limit~\cite{Ballesteros:2012kv}. It is straightforward to check that~\eqref{fzeta} can be obtained from~\eqref{eqz} in the case of an ``adiabatic'' solid, i.e.\ one that satisfies $c_L^2=c_a^2$. Also, if this adiabaticity condition holds, the equation~\eqref{eqR} for $\mathcal{R}$ becomes
\begin{align}
\ddot{\mathcal{R}}&+(3-2s_L+\epsilon_2)H\dot{\mathcal{R}}+\frac{c_L^2}{a^2}k^2\mathcal{R}=0\;.
\end{align}
Assuming the generalized slow-roll approximation, we see from these equations that $\zeta$ and $\mathcal{R}$ for an effective perfect fluid are equal to each other up to a sign (and constant) for $k\ll H$ up to corrections that are of first order in (generalized) slow-roll. Therefore, once inflation has ended there is a single degree of freedom $\zeta\sim \mathcal{R}$ that controls PBH formation when $k\sim \mathcal{H}$.

In reference \cite{Endlich:2012pz} it is assumed that the transition from inflation to radiation domination (the reheating process) occurs so fast that to all effects can be considered instantaneous. The energy density of the solid is entirely transferred into radiation density, but since this happens instantaneously and the pressure of the inflationary solid is different from that of the radiation fluid, a discontinuity in $\dot H$ appears  \cite{Endlich:2012pz}. Using  \eq{E1z} we can express $\mathcal{R}$ in the solid phase as a function of $\zeta$ and its time derivative: 
\begin{align}
\mathcal{R}=\frac{\dot H\,\zeta-\dot\zeta\,H}{\dot H-k^2/(3a^2)}\,.
\end{align}
Then, assuming that $\zeta$ transitions into the fluid phase smoothly --which is argued in \cite{Endlich:2012pz} to be a good approximation for generalized slow-roll-- this expression and the assumption of instantaneous reheating imply that $\mathcal{R}$ becomes discontinuous at the transition due to the behaviour of $\dot H$. This is used in \cite{Endlich:2012pz} to conclude that the power spectrum of $\zeta$ right before reheating is a good approximation to the primordial spectrum inherited in the post-inflationary phase. Following this reasoning and using $\eq{xacte}$ we finally get 
\begin{align} \label{fpbhsi}
\Delta_\delta^2\simeq 8 \Delta_\zeta^2\simeq  \frac{H^2}{\pi^2  \epsilon c_L^5 M_P^2}.
\end{align}
Therefore, with the above assumptions, the enhancement of the density contrast (in the TMG) for small $c_L$ goes as $\sim 1/c_s^2$ and does appear as a particulary interesting feature of solid inflation for PBH formation. However, it has to be remarked that the discontinuity in $\dot H$ is unphysical and arises just as a consequence of  assuming that reheating is instantaneous. A detailed analysis of the reheating process is in principle required to determine how the physics of solid inflation gets imprinted in the radiation density contrast and thus into the formation of PBHs.\footnote{The relevance of reheating for the predictions of solid inflation has also been highlighted recently in \cite{Bordin:2017ozj}.}

Even if we assume the validity of \eq{fpbhsi}, there is a subtle (but important point) that needs to be looked at before naively using that expression for actual numerical estimates of PBH formation in solid inflation. The model is a derivatively coupled EFT, which therefore has an ultraviolet cut-off. Similarly to what happens with the standard EFT of inflation, this leads to a restriction on the smallness of the parameters ($\epsilon$ and $c_L$) involved in determining the amplitude of primordial perturbations.  Since the interaction terms in the Lagrangian for fluctuations appear divided by increasingly higher powers of $c_L$ as the number of derivatives grows, the cut-off must be proportional to a positive power of $c_L$. Small values of $c_L$ (such as the ones expected of being capable of triggering PBH formation) imply a relatively low cut-off and we have to check whether the validity of the EFT becomes compromised. An order of magnitude estimate of the ultraviolet cut-off was done in~\cite{Endlich:2012pz} by comparing the size of the cubic and quadratic terms for fluctuations. Imposing that $H$ (which is the main scale in the problem) has to be smaller than the cut-off, one obtains that the condition 
\begin{align} \label{condp}
\epsilon^3 c_L^9 \gg \frac{H^2}{M_P^2}
\end{align}
has to be satisfied  for small $c_L^2$  to keep the validity of the perturbative expansion. Together with the expression~\eqref{fpbhsi}, this gives
\begin{align} \label{eftbounds}
\left(\epsilon c_L^2\right)^2\gg 8\pi^2\Delta_\zeta^2\;.
\end{align}
At the scales probed by the CMB there is no problem in applying the EFT since the right-hand side of the last inequality is very small: $\mathcal{O}(10^{-7})$. The tensor-to-scalar ratio in solid inflation is: 
\begin{align}
r\simeq 4\epsilon c_L^5\,.
\end{align} 
This simply says that a large tensor-to-scalar ratio (compatible with the current bound) requires $c_L^2\lesssim 1$ in the regime of validity of the EFT, which is fine. However, 
imposing
\begin{align} \label{estimate}
\Delta_\zeta^2\sim 10^{-2}\;,
\end{align}
which as we already discussed is thought to be needed (assuming Gaussian perturbations) to account for all DM, we obtain that 
\begin{align}
\epsilon c_L^2\gg 1\;.
\end{align}
This is not allowed because we are assuming slow-roll.\footnote{Also, we want to avoid superluminality. Notice that superluminality is not a problem on its own, but it prevents the existence of a  Lorentz invariant ultraviolet completion, see~\cite{Endlich:2012pz} and~\cite{Adams:2006sv}; and, in any case, we are interested in the small $c_L$ regime.} This result suggests that the EFT fails for the values of $\Delta_\zeta^2$ that are relevant for significant PBH formation and obtaining a large DM abundance. We can make a quick numerical estimate of how small $\Delta_\zeta^2$ has to be in order to comply with the slow-roll approximation and the EFT range for a moderately small $c_L$. For instance, if we take $\epsilon\simeq 0.05$, the expression~\eqref{condp} tells that $\Delta_\zeta^2$ has to be at most $\mathcal{O}(10^{-6})$  for a $c_L^2\sim 0.1$. This is a tiny amplitude in comparison to $\mathcal{O}(10^{-2})$, specially taking into account that the PBH abundance is exponentially sensitive to it. Indeed, we can write an inequality similar to the one we obtained for the EFT of inflation, equation \eq{cmbb}:
\begin{align}
\epsilon_{\rm PBH}\gg  4\times 10^{5} \left(\frac{0.07}{r_{\rm CMB}}\right)\left(\frac{\Delta_{\zeta\,{\rm CMB}}^2}{2\times 10^{-9}}\right)^{-2/3}\left(\frac{\Delta_{\zeta\,{\rm PBH}}^2}{0.01}\right)^{3/2}\,.
\end{align}
Whereas in the case of EFT of inflation the speed of sound squared has to be much larger than the square root of $\Delta_\zeta^2$ for consistency --see equation \eq{eftbound}--, in solid inflation the value of $\epsilon$ at PBH scales has to take a very large value.

As we discussed earlier in the context of the EFT of inflation, before ditching completely  the hope of obtaining interesting and numerically sensible results for PBH formation in solid inflation, it has to be noted that the approximation of Gaussianity for the primordial fluctuations is most likely not applicable. Non-Gaussianities are actually one of the most interesting aspects of solid inflation and their strength grows for small values of $c_L$. In fact, they can be remarkably large, with $f_{\text{NL}}\sim 1/(\epsilon\, c_L^2)$~\cite{Endlich:2012pz}, which is the very reason why the EFT is in trouble for very small $c_L^2$. As we have already mentioned, large non-Gaussianities can affect importantly PBH formation and it is possible that a smaller value of $\Delta_\zeta^2$ than $\mathcal{O}(0.01)$ could be sufficient to generate a significant PBH abundance. In order to assess accurately the relevance of this point, a detailed analysis is necessary, because the effect depends on the specific non-Gaussian shape and, in principle, could also tend to disfavor PBH formation. Indeed, in solid inflation non-Gaussianities peak for squeezed Fourier triangles, which also happens for the ``local'' non-Gaussianity generated in the ultra slow-roll case of a single field, which has been argued to {\it oppose} PBH formation~\cite{Franciolini:2018vbk}. However, since the non-Gaussianities of solid inflation are also sizable for other Fourier configurations, such as the orthogonal one, we cannot translate that result directly to this case. Moreover, given the exponential sensitivity of the PBH abundance to $\Delta_\zeta^2$  --inherent to the Press-Schechter formalism-- it is difficult to estimate an order of magnitude for the abundance without a fully tailored analysis.

To summarize, solid inflation has several distinctive features that make it interesting in the context of PBH formation: the super-horizon evolution of gauge invariant fluctuations, a possibly very strong dependence of the radiation density on the speed of sound of longitudinal modes, $c_L$, and large non-Gaussianities that are enhanced for small values of $c_L$. Further work exploring these aspects to extract numerically sound results is definitely worthwhile.

\section{Conclusions} \label{Conc}
PBHs are currently a contending DM candidate in the region of low-masses that goes from $10^{-16} M_\odot$ to $10^{-14} M_\odot$ and possibly also from $10^{-13} M_\odot$ to $10^{-10.5} M_\odot$. They are also potentially interesting as a probe of primordial physics at small distance scales that are inaccessible through the CMB. Moreover, the LIGO/Virgo detections of heavy BH mergers raises the obvious and yet unsettled question of their origin; whose full answer might not lie in a standard astrophysical processes. For all these reasons, elucidating the possible mechanisms of PBH formation and their implications for cosmology, astrophysics and high-energy physics is a timely matter. 

PBHs may have formed in the early Universe when large local overdensities collapse due to their gravitational instability. In this paper we have focused on the case in which these overdensities are sourced by scalar fluctuations generated during inflation. In order for PBH production to be efficient, the amplitude of the scalar fluctuations has to be much larger than the one inferred at CMB scales, implying that a strong enhancement of the scalar power spectrum at smaller scales is required, typically 30 to 40 e-folds away from the CMB scales. Since in standard slow-roll inflation the scalar power spectrum is inversely proportional to the slow-roll parameter $\epsilon$, most studies have focused on an approximate inflection point in the inflationary potential, which has the effect of slowing down the classical field evolution. 

In this work we have taken a broader approach by considering a general quadratic action for primordial fluctuations, which encompasses a wide variety of inflationary models. Aside from $\epsilon$, the action depends on other functions of time, whose variation may also lead to PBH formation. Concretely, we have considered the following action for the comoving curvature perturbation $\mathcal{R}$:
\begin{equation} 
\mS=  \int \dd t \, \dd^3 x \, M^2\frac{a^3 \epsilon}{\cs^2} \left[ \dot{\mR}^2 - \frac{\cs^2}{a^2 } \vert {\vec{\nabla}} \mR\vert^2 - m^2\mR^2 \right]\,,
\label{Eq:quadraticgeneral2copy}
\end{equation}
which contains the speed of propagation of the fluctuations, $\cs$, an effective Planck mass, $M$, and also an effective mass, $m$, for the fluctuations themselves. As explained in Section \ref{genactionM}, each of these functions harbingers different kinds of physics during inflation, which could then be eventually probed through PBH detection.  

A varying $\cs$ generically arises if high order derivative terms are present in the action for the inflaton, whereas a time-dependent $M$ implies the existence of certain non-minimal couplings between the inflaton and the Riemann tensor. Notably, $\epsilon$ and $M$ are indistinguishable at the level of the action \eq{Eq:quadraticgeneral2copy}, but this degeneracy is broken by the background evolution. Finally, $m^2$ is absent in standard models for which $\mathcal{R}$ is conserved in the small $k$ limit.

We have analyzed the possible dynamics encoded in \eq{Eq:quadraticgeneral2copy} that can lead to an enhancement of the primordial spectrum, focusing mainly on the case $m^2=0$. Each Fourier mode is a linear combination of two different solutions, which we have characterized in detail using the WKB approximation. First, we have provided an expression, equation \eq{eq:PSsr}, which generalizes the standard slow-roll approximation for the usually dominant (and constant) solution of $\mathcal{R}$. Then, we have considered the possibility --of practical relevance for PBH formation-- in which a bunch of Fourier modes undergoes several ``horizon'' crossings. By carefully matching the solution at each crossing, we are able to obtain an analytic expression which reproduces the main features of the spectrum and signals the origin of oscillations. We have also explored the conditions under which the usually decaying solution for $\mathcal{R}$ can become the dominant one. This occurs if a combination of slow-roll parameters, the friction parameter of equation \eq{frictious}, becomes negative. We have also written a general expression for the full solution, which allows to obtain recursively approximations of increasing precision to actual numerical result for $\mathcal{R}$.

We have illustrated the different effect on the spectrum of primordial fluctuations with a phenomenological parametrization of the relevant functions of time. For example, we show how by synchronizing their variations its possible to produce a large enhancement from mild individual changes. Conversely, by separating them in time it is possible to get PBH production at different mass scales. We have also given a concrete toy model, based on DBI inflation, which can develop some of the key features on the evolution of $\mathcal{R}$ for adequate values of its parameters. Specifically, this example shows how the ``decaying'' mode can become the dominant one and how oscillations in the spectrum arise from multiple ``horizon'' crossings. 

In the context of the {(slow-roll)} EFT of inflation of \cite{Cheung:2007st}, a small $\cs$ implies large interactions and a reduction of the strong coupling scale. We argue that this impedes a consistent calculation for large enhancements of the primordial amplitude. However, we point out that a modified dispersion relation, which arises in models such as the ghost condensate may allow to circumvent this conclusion, still within a consistent EFT perspective. 

We have considered as well PBH formation in solid inflation, where again the consistency of the EFT clashes with a large amplification of the primordial fluctuations. This model is an example in which the scalar fluctuations evolve in the small $k$ limit. In this respect, this provides an example of a model with $m^2\neq 0$.  Moreover, the primordial gauge invariant variables $\mathcal{R}$ and $\zeta$ --see equations \eq{Req} and \eq{defzeta}-- differ from each other in this regime by a high power of the speed of propagation of longitudinal modes. It is then not obvious a priori which of the two should be used to describe PBH formation. The reheating process in solid inflation becomes relevant in this context, not only due to the non-constancy of the fluctuations for small $k$. but also because the standard assumption that is made to describe it \cite{Endlich:2012pz} singles out $\mathcal{\zeta}$ as the key variable. 

We have also discussed in detail how these variables,  $\zeta$ and $\mathcal{R}$,  are related to the energy density contrast --see equation \eq{xacte}-- in the appropriate gauge to describe the collapse, which we argue is the total matter (or orthogonal comoving) one, among other reasons because it leads to a well defined variance of the smoothed radiation density contrast.

We have stressed throughout the paper that most of our assumption of Gaussian primordial fluctuations. Including non-Gaussianities and, also, the effects of quantum diffusion \cite{Pattison:2017mbe,Biagetti:2018pjj,Ezquiaga:2018gbw,Cruces:2018cvq} are subsequent steps worth exploring. 

Our work can be the basis for the construction of field theoretic models of inflation --beyond a minimally coupled single scalar-- able to produce large abundances of PBHs from an enhancement of the primordial spectrum. Part of our analysis can also be used to describe smaller deviations from near-scale invariance that might be probed via CMB and LSS studies.

\vspace{0.5cm}
\subsubsection*{Acknowledgements}

We would like to thank Juan García-Bellido for collaboration in the early stages of this project and useful discussions. The work of GB is funded by a {\it Contrato de Atracci\'on de Talento (Modalidad 1) de la Comunidad de Madrid} (Spain), with number 2017-T1/TIC-5520. It has also been supported by {\it MINECO} (Spain) under contract FPA2016-78022-P. JBJ acknowledges support from the  {\it Atracci\'on del Talento Cient\'ifico en Salamanca} programme and the MINECO's projects
FIS2014-52837-P and FIS2016-78859-P (AEI/FEDER). The research of MP  is funded by EU's Horizon 2020 through the  {\it InterTalentum} programme (Project ID: 713366). We acknowledge the support of the Spanish MINECO's {\it Centro de Excelencia Severo Ochoa Program} under the grants SEV-2012-0249 and SEV-2016-0597. GB thanks the hospitality of the CERN Theory Department while part of this work was developed.
	
\bibliographystyle{hunsrt}
\bibliography{Speed_of_sound.bib}

\end{document}